\documentclass[final,10pt,twocolumn]{IEEEtran}
\usepackage{amsmath,amssymb,graphicx,psfrag,cite,url}

\usepackage[caption=false,font=footnotesize]{subfig}
\usepackage{color}
\newcommand{\bc}{\begin{center}}
\newcommand{\ec}{\end{center}}

\newcommand{\ba}{\begin{array}}
\newcommand{\ea}{\end{array}}

\newcommand{\bt}{\begin{tabular}}
\newcommand{\et}{\end{tabular}}

\newcommand{\eems}{\end{multline*}}
\newcommand{\eem}{\end{multline}}
\newcommand{\eeams}{\end{eqnarray}\vspace{0.1cm}}
\newcommand{\eearms}{\end{eqnarray*}\vspace{0.1cm}}
\newcommand{\edmms}{\end{displaymath}\vspace{0.1cm}}

\newcommand{\bems}{\begin{multline*}}
\newcommand{\bem}{\begin{multline}}
\newcommand{\beams}{\vspace{0.1cm} \begin{eqnarray}}
\newcommand{\bearms}{\vspace{0.1cm} \begin{eqnarray*}}
\newcommand{\bdmms}{\vspace{0.1cm} \begin{displaymath}}

\newcommand{\be}{\begin{equation}}
\newcommand{\ee}{\end{equation}}

\newcommand{\bes}{\begin{equation*}}
\newcommand{\ees}{\end{equation*}}

\newcommand{\bea}{\begin{eqnarray}}
\newcommand{\eea}{\end{eqnarray}}

\newcommand{\bear}{\begin{eqnarray*}}
\newcommand{\eear}{\end{eqnarray*}}

\newcommand{\bdm}{\begin{displaymath}}
\newcommand{\edm}{\end{displaymath}}

\newcommand{\feop}{\hfill \rule{1.5mm}{1.5mm} \\}

\newcommand{\bit}{\begin{itemize}}
\newcommand{\eit}{\end{itemize}}

\newcommand{\ben}{\begin{enumerate}}
\newcommand{\een}{\end{enumerate}}

\newcommand{\bd}{\begin{document}}
\newcommand{\ed}{\end{document}}

\newcommand{\psp}{\hspace{0.6cm}}
\newcommand{\hspone}{\hspace{1cm}}
\newcommand{\hsptwo}{\hspace{2cm}}

\newcommand{\hspthree}{\hspace{3cm}}
\newcommand{\hspfour}{\hspace{4cm}}

\newcommand{\hspfive}{\hspace{5cm}}
\newcommand{\hspsix}{\hspace{6cm}}

\newcommand{\hspseven}{\hspace{7cm}}
\newcommand{\hspeight}{\hspace{8cm}}

\newcommand{\vspone}{\vspace{1cm}}
\newcommand{\vsptwo}{\vspace{2cm}}

\newcommand{\vspthree}{\vspace{3cm}}
\newcommand{\vspfour}{\vspace{4cm}}

\newcommand{\vspfive}{\vspace{5cm}}
\newcommand{\vspsix}{\vspace{6cm}}

\newcommand{\vspseven}{\vspace{7cm}}
\newcommand{\vspeight}{\vspace{8cm}}

\newcommand{\vspnine}{\vspace{9cm}}
\newcommand{\vspten}{\vspace{10cm}}

\newcommand{\vspeleven}{\vspace{11cm}}
\newcommand{\vsptwelve}{\vspace{12cm}}

\newcommand{\qett}{\mbox{$(q^{-1})$}}
\newcommand{\zett}{\mbox{$(z^{-1})$}}

\newcommand{\qo}{q^{-1}}
\newcommand{\zettu}{\mbox{$z^{-1}$}}

\newcommand{\linf}[1]{{#1}\rightarrow \infty}

\newcommand{\cf}[1]{(\ref{#1})}

\newcommand{\eiw} {\mathrm{e}^{\mathrm{i}\omega}}
\newcommand{\emiw}{\mathrm{e}^{-\mathrm{i}\omega}}

\newcommand{\tr}{\mbox{${\rm Tr}$}}
\newcommand{\var}{\mbox{${\rm Var}$}}
\newcommand{\diag}{\mbox{${\rm Diag}$}}
\newcommand{\cov}{\mbox{${\rm Cov}$}}
\newcommand{\Ex}{{\sf E}}

\newcommand{\erm}{\mathrm{e}}
\newcommand{\irm}{\mathrm{i}}
\newcommand{\Prj} {{\bf \Pi}^\bot}

\newcommand{\bA} {{\bf A}}
\newcommand{\bB} {{\bf B}}
\newcommand{\bC} {{\bf C}}
\newcommand{\bD} {{\bf D}}
\newcommand{\bE} {{\bf E}}
\newcommand{\bF} {{\bf F}}
\newcommand{\bG} {{\bf G}}
\newcommand{\bH} {{\bf H}}
\newcommand{\bI} {{\bf I}}
\newcommand{\bJ} {{\bf J}}
\newcommand{\bK} {{\bf K}}
\newcommand{\bL} {{\bf L}}
\newcommand{\bM} {{\bf M}}
\newcommand{\bN} {{\bf N}}
\newcommand{\bO} {{\bf O}}
\newcommand{\bP} {{\bf P}}
\newcommand{\bQ} {{\bf Q}}
\newcommand{\bR} {{\bf R}}
\newcommand{\bS} {{\bf S}}
\newcommand{\bT} {{\bf T}}
\newcommand{\bU} {{\bf U}}
\newcommand{\bV} {{\bf V}}
\newcommand{\bW} {{\bf W}}
\newcommand{\bY} {{\bf Y}}
\newcommand{\bX} {{\bf X}}
\newcommand{\bZ} {{\bf Z}}

\newcommand{\GLm} {{\bf \Lambda}}
\newcommand{\GSg} {{\bf \Sigma}}
\newcommand{\GPh} {{\bf \Phi}}
\newcommand{\GPs} {{\bf \Psi}}
\newcommand{\GPi} {{\bf \Pi}}
\newcommand{\GDl} {{\bf \Delta}}
\newcommand{\GGm} {{\bf \Gamma}}
\newcommand{\GUp} {{\bf \Upsilon}}
\newcommand{\GOm} {{\bf \Omega}}
\newcommand{\GTh} {{\bf \Theta}}
\newcommand{\GXi} {{\bf \Xi}}

\newcommand{\gal} {\boldsymbol{\alpha}}
\newcommand{\gbt} {\boldsymbol{\beta}}
\newcommand{\ggm} {\boldsymbol{\gamma}}
\newcommand{\gdl} {\boldsymbol{\delta}}
\newcommand{\gep} {\boldsymbol{\epsilon}}
\newcommand{\vep} {\boldsymbol{\varepsilon}}
\newcommand{\gzt} {\boldsymbol{\zeta}}
\newcommand{\get} {\boldsymbol{\eta}}
\newcommand{\gth} {\boldsymbol{\theta}}
\newcommand{\vth} {\boldsymbol{\vartheta}}
\newcommand{\git} {\boldsymbol{\iota}}
\newcommand{\gkp} {\boldsymbol{\kappa}}
\newcommand{\glm} {\boldsymbol{\lambda}}
\newcommand{\gmu} {\boldsymbol{\mu}}
\newcommand{\gnu} {\boldsymbol{\nu}}
\newcommand{\gxi} {\boldsymbol{\xi}}
\newcommand{\gpi} {\boldsymbol{\pi}}
\newcommand{\vpi} {\boldsymbol{\varpi}}
\newcommand{\grh} {\boldsymbol{\rho}}
\newcommand{\vrh} {\boldsymbol{\varrho}}
\newcommand{\gsg} {\boldsymbol{\sigma}}
\newcommand{\vsg} {\boldsymbol{\varsigma}}
\newcommand{\gtu} {\boldsymbol{\tau}}
\newcommand{\gup} {\boldsymbol{\upsilon}}
\newcommand{\gph} {\boldsymbol{\phi}}
\newcommand{\vph} {\boldsymbol{\varphi}}
\newcommand{\gch} {\boldsymbol{\chi}}
\newcommand{\gps} {\boldsymbol{\psi}}
\newcommand{\gom} {\boldsymbol{\omega}}

\newcommand{\Ba}  {{\bf a}}
\newcommand{\Bb}  {{\bf b}}
\newcommand{\Bc}  {{\bf c}}
\newcommand{\Bd}  {{\bf d}}
\newcommand{\Be}  {{\bf e}}
\newcommand{\Bf}  {{\bf f}}
\newcommand{\Bg}  {{\bf g}}
\newcommand{\Bh}  {{\bf h}}
\newcommand{\Bi}  {{\bf i}}
\newcommand{\Bj}  {{\bf j}}
\newcommand{\Bk}  {{\bf k}}
\newcommand{\Bl}  {{\bf l}}
\newcommand{\Bm}  {{\bf m}}
\newcommand{\Bn}  {{\bf n}}
\newcommand{\Bo}  {{\bf o}}
\newcommand{\Bp}  {{\bf p}}
\newcommand{\Bq}  {{\bf q}}
\newcommand{\Br}  {{\bf r}}
\newcommand{\Bs}  {{\bf s}}
\newcommand{\Bt}  {{\bf t}}
\newcommand{\Bu}  {{\bf u}}
\newcommand{\Bv}  {{\bf v}}
\newcommand{\Bw}  {{\bf w}}
\newcommand{\Bx}  {{\bf x}}
\newcommand{\By}  {{\bf y}}
\newcommand{\Bz}  {{\bf z}}

\newcommand{\za} {\boldsymbol{a}}
\newcommand{\zb} {\boldsymbol{b}}
\newcommand{\zc} {\boldsymbol{c}}
\newcommand{\zd} {\boldsymbol{d}}
\newcommand{\ze} {\boldsymbol{e}}
\newcommand{\zf} {\boldsymbol{f}}
\newcommand{\zg} {\boldsymbol{g}}
\newcommand{\zh} {\boldsymbol{h}}
\newcommand{\zi} {\boldsymbol{i}}
\newcommand{\zj} {\boldsymbol{j}}
\newcommand{\zk} {\boldsymbol{k}}
\newcommand{\zl} {\boldsymbol{\ell}}
\newcommand{\zm} {\boldsymbol{m}}
\newcommand{\zn} {\boldsymbol{n}}
\newcommand{\zo} {\boldsymbol{o}}
\newcommand{\zp} {\boldsymbol{p}}
\newcommand{\zq} {\boldsymbol{q}}
\newcommand{\zr} {\boldsymbol{r}}
\newcommand{\zs} {\boldsymbol{s}}
\newcommand{\zt} {\boldsymbol{t}}
\newcommand{\zu} {\boldsymbol{u}}
\newcommand{\zv} {\boldsymbol{v}}
\newcommand{\zw} {\boldsymbol{w}}
\newcommand{\zx} {\boldsymbol{x}}
\newcommand{\zy} {\boldsymbol{y}}
\newcommand{\zz} {\boldsymbol{z}}

\newcommand{\hide}[1] {}

\newtheorem{Thm}{Theorem}
\newtheorem{Pro}{Proposition}
\newtheorem{Defn}{Definition}
\newtheorem{Exm}{Example}
\newtheorem{Lem}{Lemma}
\newtheorem{Asm}{Assumption}
\newtheorem{Cor}{Corollary}
\newtheorem{Alg}{Algorithm}
\newtheorem{paRem}{Remark}
\newenvironment{Rem}{\begin{paRem}\rm }{\end{paRem}}

\newcommand{\ColvecTwo}[2]{\left[\ba{c} {#1} \\ {#2} \ea \right]}
\newcommand{\colvecTwo}[2]{\left[\ba{cc} {#1} & {#2} \ea \right]^\top}

\newcommand{\colvecThr}[2]{\left[\ba{ccc} {#1} & \cdots & {#2} \ea \right]^\top}
\newcommand{\ColvecThr}[2]{\left[\ba{c} {#1} \\ \vdots \\ {#2} \ea \right]}
\newcommand{\rowvecThr}[2]{\left[\ba{ccc} {#1} & \cdots & {#2} \ea \right]}
\newcommand{\MatThrTwo}[4]{\left[\ba{ccc} {#1}  & {#3} \\ \vdots & \vdots \\ {#2} & {#4} \ea \right]}

\date{}
\newcommand{\nN}{G.N_1}
\newcommand{\Set}{\mathcal{I}}
\newcommand{\mI}{\mathcal{J}}

\title{Zadoff-Chu sequence design for random access initial uplink synchronization}
\author{Md Mashud Hyder and Kaushik Mahata 
\thanks{Department of Electrical Engineering, The University of Newcastle, Australia.}
}

\begin{document}
\maketitle

\begin{abstract} The autocorrelation of a Zadoff-Chu (ZC) sequence with a non-zero cyclically shifted version of itself is zero. Due to the interesting property, ZC sequences are widely used in the LTE air interface in the primary synchronization signal (PSS), random access preamble (PRACH), uplink control channel (PUCCH) etc. However, this interesting property of ZC sequence is not useful in the random access initial uplink synchronization problem due to some specific structures of the underlying problem. In particular, the state of the art uplink synchronization algorithms do not perform equally for all ZC sequences. In this work, we show a systematic procedure to choose the ZC sequences that yield the optimum performance of the uplink synchronization algorithms. At first, we show that the uplink synchronization is a sparse signal recovery problem on an overcomplete basis. Next, we use the theory of sparse recovery algorithms and identify a factor that controls performance of the algorithms. We then suggest a ZC sequence design procedure to optimally choose this factor. The simulation results show that the performance of most of the state of the art uplink synchronization algorithms improve significantly when the ZC sequences are chosen by using the proposed technique. 

%%\boldmath
%Random access (RA) uplink synchronization is a contention-based synchronization procedure in the LTE system that allows the 
%eNodeB to detect the user equipments (UEs) that are willing to 
%commence communication. It also enables the eNodeB to estimate the uplink channel parameters of those UEs.
%Accurate estimation of these parameters are crucial as they
%ensure that the uplink signals from all the UE arrive at the eNodeB synchronously 
%and approximately at the same power level. In this work, we apply sparse signal recovery methods to resolve the problem. We show that the RA uplink signal exhibits some interesting properties that allow us to apply some efficient sparse recovery methods for resolving the RA uplink synchronization problem. We further develop a procedure to generate RA code matrix (i.e, RA preamble) which enhances the performance of sparse recovery algorithm and helps us to develop time efficient algorithms. We observe that the proposed code matrix  can increase the performance of state of the art uplink synchronization algorithms as well. 
\end{abstract}
\begin{IEEEkeywords}
Random Access, initial synchronization, Zadoff-Chu sequence, sparse representation, OFDMA.
\end{IEEEkeywords}

\section{Introduction}
\subsection{Background}
Many wireless communication systems adopt orthogonal frequency-division multiple access (OFDMA) technology for data transmission such as LTE \cite{lte4}, WiMAX \cite{rang8} etc. To maintain orthogonality among the subcarriers in the uplink of OFDMA system, the uplink signals arriving at the eNodeB from different user equipments (UEs) should be aligned with the local time and frequency references. For this purpose, all UEs who want to set up connection to the eNodeB must go through an initial uplink synchronization (IUS) procedure.
The IUS allows the eNodeB to detect these users, and estimate their channel parameters. These estimates are then used to time-synchronize UEs' transmissions and
adjust their transmission power levels so that all uplink signals arrive at the eNodeB synchronously and at approximately the same power level \cite{rang3}.  
The initial uplink synchronization  is a contention based random access (RA) process. The process starts with the allocation of a pre-defined set of 
subcarriers called Physical Random Access Channel (PRACH) in some pre-specified time slots known as a ``random access opportunity"  \cite{lte4}. 
%This is called a ``RA opportunity''.
The downlink synchronized UEs willing to commence communication, 
referred to as the synchronizing random access terminals (RTs), use this opportunity 
by modulating randomly selected codes (also called RA preambles) from a pre-specified RA code matrix onto the PRACH. 
As the RTs are located in different positions within the radio coverage
area, their signals arrive at the eNodeB with different time 
delays. At the receiving side, the eNodeB needs to detect the transmitted 
RA codes, and extract the timing and channel power
information for each detected code \cite{rang1,rang5,rang3,lte2}.

The success of LTE in 4G cellular networks has made it popular. It is also envisaged that many of its key ingredients will also dominate the 5G systems. In this work, we consider the uplink data transmission protocol of an LTE-like system \cite{lte2}.
Zadoff-Chu sequences \cite{zc1,zc10} are used as random multiple access codes in the LTE system due to their perfect autocorrelation properties i.e., the autocorrelation of a ZC sequence with a non-zero cyclically shifted version of itself is zero. However, the eNodeB cannot fully exploit the perfect autocorrelation property of ZC sequences in different types of synchronization problems due to some factors. Hence, designing appropriate ZC sequences for different types of synchronization problems have received lots of research interests. A training-aided ZC sequence design procedure has been proposed in \cite{zc9} for the frequency synchronization problem. The frequency synchronization deals with the estimation and compensation of carrier frequency offset (CFO) between the transmitter and a receiver. The CFO arises mainly due to the Doppler shift in downlink synchronization \cite{rang3,zc9}.
%has received a lot of research interest. 
%Recently, some ZC sequences design methods have been proposed which can be used for the estimation of  timing and frequency offsets of UEs. 
The effect of carrier frequency offset on the autocorrelation property of the ZC sequences has been addressed in \cite{zc8,zc5}. A closed-form expression of  the autocorrelation between two cyclically shifted ZC sequences in presence of frequency offset has been developed in \cite{zc5}. Consequently, an empirical method has been proposed to design two different sets of ZC sequences: one for  high frequency offset scenario and other for low frequency offset scenario, which can be used for the frequency offset estimation of UEs. The ZC sequence design for the timing synchronization problem has been addressed in \cite{zc9,zc6,zc7}. The timing synchronization deals with the estimation of timing offset which arises due to the random propagation delay between eNodeB and user \cite{zc9}.
%However, the analysis assumes that there exists only one channel tap between each user and eNodeB. 
%The analysis will be difficult for the channels subject to multipath dispersion. 
The method developed in \cite{zc6} transmits partial Zadoff-Chu sequences on disjoint sets of equally spaced subcarriers for timing offset estimation. The signature sequences design procedure proposed in \cite{zc7} considers the time and frequency synchronization in presence of carrier frequency offset between the eNodeB and single UE. To the best of our knowledge, the ZC sequence design for the IUS problem has not been addressed before. The IUS problem is different from the frequency and timing synchronization problems. In the IUS, the eNodeB has to detect multiple RTs as well as extract their channel power and timing offset information. The effects of multiple-access interference (MAI) and unknown multipath channel impulse responses of RTs make the ZC sequence design problem challenging. It this work, we consider the ZC sequence design problem for the IUS purpose.

\subsection{Some motivating examples}
In this section, we observe that the performance of state of the art initial uplink synchronization (IUS) algorithms \cite{rang1,rang5,rang3,lte2} can vary significantly depending on the RA code matrix. Moreover, the IUS algorithms  may not yield the optimum performance with the RA code matrix generated by using the conventional procedure. To explain the matter, we first describe the procedure of RA code generation \cite{lte4,lte9}. 

To give example we consider the LTE system proposed in \cite{lte2} where $N=6144$. Total $M=839$ number of adjacent subcarriers are allocated for the PRACH. The RA codes for the IUS are generated by cyclically shifting the Zadoff-Chu (ZC) sequence \cite{zc1,zc10}. The elements of the $u$-th root ZC sequence are given by
\begin{align}\label{eq:zc1}
Z^{u}(k)=\erm^{-\irm \pi u k(k+1)/M}, \ \ 0\le k< M
\end{align}
where ${u}$ is a positive integer with $u<M$. Different RA codes are obtained by cyclically shifting the $u$-th root ZC sequence. Let $\zc_{\ell}^{(u)}$ be the $\ell$-th RA code. Then the $(k+1)$-th element of $\zc_{\ell+1}^{(u)}$ is given by 
\begin{align}\label{eq:zc2}
\zc^{(u)}_{k+1,\ell+1}=Z^u\left\{(k+\ell\ n_{cs})\mod M\right\}, \ 0\le k < M.
\end{align}
Here  $(\ell \ \mathrm{mod} \ M) := \ell - M \cdot \lfloor \ell/M 
\rfloor $, with $\lfloor r \rfloor$ denoting the largest integer less than or equal to $r$. In addition, $n_{cs}$ is an integer valued system parameter which is related to the wireless cell radius \cite[eq. (17.10)]{zc8}:
\begin{align}\label{eq:ncs}
n_{cs}\ge \Big\lceil \left(\frac{20}{3}\gamma-\tau_d\right)\frac{M}{T_{SEQ}}\Big\rceil+n_g,
\end{align}
where $\lceil z \rceil$ is the smallest integer not less than $z$, $\gamma$ is the cell radius (km), $\tau_d$ is the maximum delay spread ($\mu$s), $T_{SEQ}$ is the preamble sequence duration ($\mu$s), and $n_g$ is the number of additional guard samples. For a wireless cell radius of $1.3$ km, typically $T_{SEQ}=800 \mu$s, and $\tau_d<1\mu$s, which yields $n_{cs}\ge 11$. The maximum number of RA codes that can be generated from a single ZC root depends on the value of $n_{cs}$. For example, setting $n_{cs}=11$, we can generate maximum $\lfloor M/n_{cs}\rfloor=76$ RA codes from the ZC root ${u}$. If we need more than $76$ RA codes then we have to utilize multiple roots. The procedure has been described explicitly in Section-\ref{tab:ra1}. 

Now, consider a RA uplink synchronization scenario where $K$ number of RTs are simultaneously contending on the same PRACH channel. The channel impulse response of the RTs have a maximum order $35$ and total $G=50$ codes are available in the RA code matrix\footnote{In practice, there are 64 RA codes available in each cell. However, some codes are reserved for contention free RACH \cite{lte9}. We assume that there are $14$ reserve codes.}:
$$\bC=[\zc_1^{(u)} \ \zc_2^{(u)} \ \cdots \ \zc_G^{(u)}].$$
We can generate different RA code matrices by using different values of $u$ and $n_{cs}\ge 11$. Table-\ref{tab:ex1} shows the user detection performances of some state of the art IUS algorithms for different code matrices\footnote{In case of $n_{cs}>15$ in Table-\ref{tab:ex1}, we need two roots to generate the code matrix.}. Note that the conventional code design procedure (see Section-\ref{tab:ra1}) suggests using $n_{cs}=11$.  However, as can be seen in Table-\ref{tab:ex1}, with $u=1$ the probability $P_s$ of detecting a given number of codes successfully by all algorithms are uniformly poor for $n_{cs}<15$. In contrast, $P_s$ for both SMUD \cite{rang1} and SRMD \cite{rang5} are above $0.9$ for $n_{cs}\ge 15$. Table-\ref{tab:ex1} also shows results for $u=2$ and $3$. However, comparing three different values of $u$, we see that the algorithms perform at their best with $u=1$ and $n_{cs}\ge 15$. 
This is interesting to note that just by taking some different combinations of $n_{cs}$ and $u$ there is a significant variation of detection performance. Moreover, this happens uniformly for both the algorithms.

The above example clearly demonstrates the importance of choosing the parameters $u$ and $n_{cs}$ carefully in order to ensure that the IUS algorithms can produce the best results. To the best of our knowledge, so far, there is no systematic study in the literature addressing the issue. In the sequel, we present a systematic procedure for finding good values of $u$ and $n_{cs}$ that give the IUS algorithms an opportunity to perform at their best.

\begin{table}
\centering \caption{IUS user detection probabilities by some algorithms for different values of $u$ and $n_{cs}$. SNR=$10$ $\zd$B, total IUS users $K=3$ and $P_s$ denotes the probability of successfully detecting the users.}
\begin{tabular} {| c | c | c | c |c|c|c|} 
\hline
 \parbox{1.1cm}{ Algorithm} &\multicolumn{6}{c |}{$P_s$ with $u=1$}\\
\cline{2-7}
  &$n_{cs}=11$ & $13$ & $14$& $15$&$17$&$19$\\
\hline\hline
 %RA-GLRT \cite{lte2}& $0.18$ & $0.15$&$0.27$&$0.3$&$0.3$&$0.31$\\
 SMUD \cite{rang1}& $0.35$ & $0.34$&$0.41$&$0.92$&$0.92$&$0.93$\\
 SRMD \cite{rang5}& $0.75$ & $0.77$&$0.79$&$0.92$&$0.94$&$0.94$\\
\hline
\hline
  &\multicolumn{3}{c |}{$P_s$ with $u=2$}&\multicolumn{3}{c |}{$P_s$ with $u=3$}\\
\cline{2-7}
 &$n_{cs}=11$ & $13$&$17$ & $11$&$13$&$17$\\
\hline\hline
%RA-GLRT \cite{lte2}& $0.06$& $0.12$ & $0.17$&$0.19$&$0.17$&$0.1$\\
 SMUD \cite{rang1}& $0.19$ & $0.3$&$0.4$&$0.77$&$0.34$&$0.29$\\
 SRMD \cite{rang5}& $0.41$ & $0.59$&$0.73$&$0.84$&$0.72$&$0.39$\\
\hline
\end{tabular}\label{tab:ex1}
\end{table}

\subsection{Contributions}
In this work, we develop a systematic procedure to study the dependency of code detection performance of the IUS algorithms on the code matrix and demonstrate an efficient code matrix design technique. The procedure can be outlined as follows. We first develop a data model of the received signal at eNodeB over the PRACH subcarriers. We represent the received signal as a linear combination of few columns of a known matrix. This matrix is constructed by using the RA codes and some sub-Fourier matrices. We further show that the data model allows us to pose the IUS parameter estimation as a sparse signal representation problem on an overcomplete basis \cite{cs2,cs4}. Thereby, sparse recovery algorithms can be used for the IUS parameter estimation problem. We then apply the compressive sensing theory\footnote{Sparse recovery algorithms and theories are generally developed in the research area called ``Compressive sensing".} to recognize a factor that controls the RA code detection performance.  The factor is called ``matrix coherence". We show that the matrix coherence of the underlying IUS problem depends on the code matrix. We then suggest a code matrix design procedure that can ensure the optimum value of  matrix coherence. The simulation results clearly demonstrate that we can indeed significantly enhance the performances of the state of the art IUS parameter estimation algorithms \cite{rang1,rang5,rang2} by using the code matrices generated by our suggested procedure. In particular, we can achieve the optimum code detection performance in Table-\ref{tab:ex1} by using our preferred code matrices.

\section{Data Model}\label{sec:data}
\subsection{Single RT}
In this work, we consider the uplink data transmission protocol of an LTE-like system \cite{lte2}. However, the proposed analysis can be extended for any LTE systems \cite{lte4}.
Consider a system with $N$ subcarriers and $N_g$ 
cyclic prefixes. Therefore, the length of an OFDM  symbol is
$\bar{N} = N + N_g$. Total $M$ adjacent subcarriers are reserved for the Physical Random Access Channel (PRACH) \cite{lte2}. We denote their indices by $\{j_m : m = 1, 2, \ldots,M\}$. When a downlink synchronized RT  wants to start communicating via the
eNodeB, it must choose a column of 
a pre-specified $M \times G$ random access (RA) code matrix
\begin{align}\label{eq:code}
\bC = [ \ \zc^{(u_1)}_1 \ \ \zc^{(u_2)}_2 \ \ \cdots \ \ \zc^{(u_G)}_G \ ]
\end{align}
uniformly at random, and send this code via the PRACH during a ``random access opportunity". Note that if all RA codes are generated from same ZC root then $u_{1}=u_{2}=\cdots \ =u_G$.
 We now mention an important property of the ZC sequence which will be helpful for developing the RA data model in the following section. One can express \eqref{eq:zc2} as:
\begin{align}\label{eq:zc3}
\zc^{(u_{\ell})}_{k+1,\ell+1}=Z^{u_{\ell}}(k)\erm^{-\irm 2\pi u_{\ell} k \ell  n_{cs}/M}\erm^{-\irm \pi u_{\ell}  \ell n_{cs} (\ell n_{cs}+1)/M}
\end{align}
with $0\le k < M$ \cite{zc1,zc2}. Since the index of $m$-th random access subcarrier is $j_m$, and the random access subcarriers are adjacent to each other, we have
$$j_{k+1}=j_1+k.$$
Substituting $k=j_{k+1}-j_1$ in \eqref{eq:zc3} gives
\begin{align}\label{eq:zc4}
\zc^{(u_{\ell})}_{k+1,\ell+1}=Z^{u_{\ell}}(k)\erm^{-\irm 2\pi u_{\ell} j_{k+1} \ell \frac{n_{cs}}{M}}\erm^{-\irm \pi u_{\ell}  \ell \frac{n_{cs}}{M} ( \ell n_{cs}+1-2j_{1})}.
\end{align}

%\section{Data Model of LTE random access}

 Let T be a specific RT which attempts to synchronize with the eNodeB at a particular RA opportunity. In particular, $T$ chooses the column $\zc_{\ell}^{(u_{\ell})}$ from $\bC$ and transmits it through PRACH to the eNodeB.  
The RT transmits $N+2N_g$ number of time domain channel symbols which are constructed from $\zc_{\ell}$. We denote
these $N+2N_g$ channel symbols by 
$\{ w(k) \}_{k=-N_g}^{N+N_g-1}$. To construct the channel symbols, at first calculate
\be
s(q) = \frac{1}{\sqrt{N}} \sum_{m=1}^M \zc_{m,\ell} \ 
\exp \{ \irm 2 \pi j_m q / N \}, \
q = \Set ,
\label{CALCULATE_S}
\ee 
where $\Set := \{ 0,1,2,\ldots,N-1 \}$,
and $\zc_{m,\ell}$ denotes the $m$ th component of $\zc_{\ell}$.
Subsequently, $\{ w(k) \}_{k=-N_g}^{N+N_g-1}$ are constructed as
\be
w(k)=\left\{
\ba{cl}
s(k \ \mathrm{mod} \ N), &-N_g\le k\le N-1,\\
0, & N\le k\le N+N_g-1,
\ea
\right.
\label{U_AND_S}
\ee
where as usual, $(k \ \mathrm{mod} \ N) := k - N \cdot \lfloor k/N 
\rfloor $. For
any  integer $k$ positive or negative, 
$ (k \ \mathrm{mod} \ N) \in \Set$. Note that the last $N_g$ symbols in \eqref{U_AND_S} are zero valued guard symbols \cite{rang11,lte2}. 
 Note that the uplink protocol of standard LTE system \cite{lte4} is slightly different than the model in \eqref{CALCULATE_S}. Standard LTE uses SC-FDMA uplink. This requires the ZC sequences first DFT pre-coded and then mapped onto subcarriers by IDFT, i.e., in \eqref{CALCULATE_S} the $\zc_{m,\ell}^{(u_{\ell})}$ will be replaced by $\hat{\zc}_{m,\ell}^{(u_{\ell})}$ where $\hat{\zc}_{m,\ell}^{(u_{\ell})}$ its the DFT of ${\zc}_{m,\ell}^{(u_{\ell})}$. Nevertheless, when the length of ZC sequence is a prime number then the DFT of ZC sequence is another ZC sequence conjugated and scaled \cite{fftzc1,fftzc2}. As a result, the proposed analysis for ZC code in OFDMA structure is fully compliant to the SC-FDMA.

Suppose $h(p), \ p \in \Set$ are the uplink channel impulse response coefficients 
between the transmitter T and eNodeB. 
Let $\{ v(k) \}_{k=-N_g}^{N+N_g-1}$ be the 
contribution of T in the symbols  
received by the eNodeB during the IUS opportunity.  
These are delayed and convoluted version of the transmitted
symbols:
\begin{align}
v(k) &=\erm^{\irm 2 \pi k \epsilon/N}\sum_{p=0}^{N-1} h(p) \ w(k-p-d), 
\ \ k \in \Set \nonumber\\
&= \erm^{\irm 2 \pi k \epsilon/N}\sum_{p=0}^{N-1} h(p) \ s \{ (k-p-d) \ \mathrm{mod} \ N \}, 
\ \ k \in \Set.
\label{CIRCULAR_CONVOLUTION}
\end{align}
%\be
%v(k) = \sum_{p=0}^{N-1} h(p) \ u(k-p-d), \ \ k \in \Set.
%\label{CONVOLUTION}
%\ee
where delay $d$ depends on the distance between T and eNodeB, and $\epsilon$ is the carrier frequency offset (CFO). During an uplink synchronization period, the CFO is mainly due to Doppler shifts and downlink synchronization error, they are assumed to be significantly smaller than the subcarrier frequency spacing. Hence, the impact of CFO on the initial uplink synchronization is generally neglected \cite{rang1,rang3,lte2}.
Thus $\{ v(k) \}_{k=0}^{N-1}$ is the result of length $N$
circular convolution of $h$, and  $s$ circularly shifted by $d$
places. The eNodeB
calculates $N$ point discrete Fourier transform (DFT) of 
$v$ to obtain
\be
V(n) = \sum_{k=0}^{N-1} v(k) \exp ( -\irm 2 \pi k n /N),
\ n \in \Set.
\label{DFT_OF_V}
\ee
Let $S$ and $H$
denote the $N$ point DFTs of $s$
and $h$, respectively. 
It is well known \cite{lte5} that 
in the DFT domain the circular convolution in \cf{CIRCULAR_CONVOLUTION} 
is equivalent to
\be
V(n) = H(n) S(n) \exp ( -\irm 2 \pi d n  /N).
\label{DFT_RELATION}
\ee
Note that equation \cf{CALCULATE_S} is compactly given in matrix form as
\be
[ \ s(0) \ \ s(1) \ \ \cdots \ \ s(N-1) \ ]^{\intercal}
= \frac{1}{\sqrt{N}}\bF^* \GTh^{\intercal} \zc_{\ell}^{(u_{\ell})},
\label{EXPRESSION_BS}
\ee
where $(.)^{\intercal}$ and $(.)^*$ denote transpose and complex conjugate transpose respectively, $\GTh$ is an $M \times N$ row selector matrix which is constructed by selecting $M$ contiguous rows from an $N \times N$ identity matrix where the first row of $\GTh$ is the $j_1$-th row of the identity matrix, and the $N \times N$ DFT matrix 
$\bF$ is defined element-wise as
\[
[ \bF ]_{k,m} = \exp \{ -  \irm 2 \pi  (k-1)(m-1) / N \} .
\]
where $[\bF]_{k,m}$ is the element of $\bF$ at its $k$-th row and $m$-th column. Thereby, the DFT of \eqref{EXPRESSION_BS} is given by
\begin{align}
&[ \ S(0) \ \ S(1) \ \ \cdots \ \ S(N-1) \ ]^{\intercal}
= \GTh^{\intercal} \zc_{\ell}^{(u_{\ell})}.
\label{S_EQN}
\end{align}
Using the relations in  \eqref{eq:zc4} and \eqref{S_EQN}, and denoting $\phi_{\ell+1}=\pi u_{\ell} \  \ell n_{cs} ( \ell n_{cs}+1-2j_{1})/M$, we can express \eqref{DFT_RELATION} as 
\be
V(j_m) = Z^{u_{\ell}}(m) H(j_m) \erm^{ -\irm \frac{2 \pi}{N} j_m(u_{\ell}(\ell-1) n_{cs}\frac{N}{M}+ d)  }\erm^{-\irm \phi_{\ell}}.
\label{DFT_RELATION2}
\ee
for $m=1,2,\cdots M$. Note that $\phi_{\ell}$ is known for all $\ell$. After calculating $V$, the eNodeB
forms the vector 
\be
\zv = [ \ V(j_1) \ \ V(j_2) \ \ \cdots \ \ V(j_M) \ ]^{\intercal},
\label{DEF_ZV}
\ee
From the theory of DFT, it is well known that $H(n)
\erm^{-\irm 2 \pi d n/N}$ is the DFT of $h$ circularly shifted 
by $d$ places \cite{lte5}, which is written compactly as
\begin{align}
\nonumber
&[ \ H(0)  \ \ H(1) \erm^{-\irm 2 \pi d/N} \ \ \cdots \ \ 
H(N-1) \erm^{-\irm 2 \pi d(N-1)/N} \ ]^{\intercal}
\\
&= \bF \zh_{\downarrow(d)},
\label{H_EQN}
\end{align}
where $\zh_{\downarrow(d)}$ is the circularly shifted version 
of the channel impulse response by $d$ places expressed as a vector:
\be
\zh_{\downarrow(d)} 
:=
[ \ h(N-d)  \cdots h(N-1) \ \ h(0)
\ \ h(1) \ \ \cdots  h(N-d-1) \ ]^{\intercal}.
\label{CIRCULAR-SHIFT}
\ee
Hence by \eqref{H_EQN} and \eqref{CIRCULAR-SHIFT}, we have
\begin{align}
&[ \ H(j_1)\erm^{-\irm \frac{2 \pi}{N}j_1 d}  \ \ H(j_2)\erm^{-\irm \frac{2 \pi}{N}j_2 d} \ \ \cdots \ \ 
H(j_M)\erm^{-\irm \frac{2 \pi}{N}j_M d} \ ]^{\intercal}\nonumber\\
&=\GTh \bF \zh_{\downarrow(d)}.
\end{align}

Hence \cf{DFT_RELATION2} and \cf{DEF_ZV} imply that
\begin{align} \label{ONE_TERMINAL_RELATION1}
\zv &= \bE_{\ell} \tilde{\zh}_{\downarrow(d)},\\
\label{eq:ell}\bE_{\ell} &=
\mathrm{diag}(Z^{u_{\ell}})\mathrm{diag}( \zp_{\ell} ) \ \GTh
\bF.\\
\tilde{\zh}_{\downarrow(d)}&=\erm^{-\irm \phi_{\ell}}{\zh}_{\downarrow(d)}\\
\zp_{\ell}&=[\erm^{ -\irm \frac{2 \pi}{N} j_1(u_{\ell}(\ell-1) n_{cs}\frac{N}{M})  }, \cdots \ , \erm^{ -\irm \frac{2 \pi}{N} j_M(u_{\ell}(\ell-1) n_{cs}\frac{N}{M})  }]
\end{align}
where diag$(\zp_{\ell})$ denotes a diagonal matrix where the vector $\zp_{\ell}$ is its diagonal entry. Typically, we  know a number $P$, known as the maximum channel order \cite{rang1,rang3}, such that
$|h(k)| = 0$ for $k \ge P$. In addition, the cell radius gives an upper bound $D$ on $d$.
Thus, by construction of 
$\tilde{\zh}_{\downarrow(d)}$, only first $D+P$ of its 
rows are non-zero. Hence it is fine to truncate $\tilde{\zh}_{\downarrow(d)}$
to a $D+P$ dimensional vector, and thus it is enough to 
work with only first $D+P$ columns of $\bE_{\ell}$. By denoting $N_1=P+D$, we can rewrite \eqref{ONE_TERMINAL_RELATION1} as
\begin{align} \label{ONE_TERMINAL_RELATION}
\zv &=\tilde{ \bE}_{\ell} \tilde{\zh}_{\downarrow(d)}(1:N_1),\\
\mathrm{where,} \ \  \tilde{\bE}_{\ell}&=\mathrm{diag}(Z^{u_{\ell}})\mathrm{diag}( \zp_{\ell} )\GTh \bF(:,1:N_1)\nonumber
\end{align}
here $\bF(:,1:N_1)$ denotes
the submatrix of $\bF$ formed by taking its first $N_1$
columns and $\tilde{\zh}_{\downarrow(d)}(1:N_1)$ denotes the sub-vector of $\tilde{\zh}_{\downarrow(d)}$ consisting
of its first $N_1$ components. Note that $\tilde{ \bE}_{\ell}$ is known to the eNodeB for any $\ell$. However, $\tilde{\zh}_{\downarrow(d)}$ is unknown. In fact, the eNodeB knows neither the values of $d$, nor $\ell$, nor the channel impulse response.

\subsection{Multiple RTs}\label{sec:mt}
Let $\tilde{N}_{\ell}$ be the number of terminals
transmitting code $\zc_{\ell}^{(u_{\ell})}$. Note that the value of $\tilde{N}_{\ell}$ can be larger than one. However,  $\tilde{N}_{\ell}>1$ implies that multiple RTs will collide by selecting the same RA code in a particular IUS opportunity. To prevent the collision and maintain $\tilde{N}_{\ell}\le 1$,
different scheduling approaches have been considered by the Third Generation 
Partnership Project \cite{lte4}. Thus, in the following, we assume $\tilde{N}_{\ell}\in\{0,1\}$.
 Suppose $d_{\ell}$ be the delay of the RT transmitting $\zc_{\ell}^{(u_{\ell})}$. Let
\be
\label{eq:hl}
\zh_{\ell} = 
\left\{
\ba{cl}
\tilde{\zh}_{\downarrow(d_{\ell})}, & \tilde{N}_{\ell} = 1,\\
0, & \tilde{N}_{\ell} = 0.
\ea
\right.
\ee
Then by the principle of superposition and using \cf{ONE_TERMINAL_RELATION} the data vector $\zy$ received by the eNodeB at the 
PRACH subchannels is given by
\begin{align}
\zy &= \bA \zx + \ze,
\label{MEASUREMENT_EQUATION}
\\
\nonumber
\zx &:= [ \ {\zh}_1^{\intercal} (1:N_1) \ \ 
{\zh}_2^{\intercal} (1:N_1) \ \ \cdots \ \ 
{\zh}_{G}^{\intercal} (1:N_1) \ ]^{\intercal},
\\
\nonumber
\bA &= [ \ \tilde{\bE}_1 \ \ \tilde{\bE}_2
\ \ \cdots \ \ \tilde{\bE}_G \ ].
\end{align}
where $\ze$ is the contribution of noise. Since $\tilde{\bE}_{\ell}$ is known for any $\ell$, $\bA$ is also known. The power received by the eNodeB 
corresponding to the code $\zc_{\ell}^{(u_{\ell})}$ is given by 
\cite[eq. (5)]{rang3}:
\begin{align}\label{eq:gk}
\Gamma_{\ell}= \frac{\sum_{m=1}^M\|\sum_{p=0}^{P-1} h_{\ell}(p)\erm^{-\irm 2 \pi p j_m/N}\|_2^2}{M}.%\tilde{\zh}_{\ell}^* \tilde{\zh}_{\ell}.
\end{align} 
where $\|\zz\|_p$ denotes the $\ell_p$ 
norm: 
$\|\zz\|_p=\left(\sum_{t}|\zz(t)|^p\right)^{1/p}$.
\subsection{IUS parameter estimation problem}
\label{sec:pro}

Given $\zy$, the eNodeB needs to
\ {\em i) find the set $\mathcal{L} = \{ \ell : \Gamma_{\ell} \ne 0 \}$;
and \ ii) for every $\ell \in \mathcal{L}$ find $\Gamma_{\ell}$ and 
$d_{\ell}$.} 

Recall that the first $d$ components of
$\Bh_{\downarrow(d)}$ are zero, see \cf{CIRCULAR-SHIFT}. 
Hence by construction of $\zh_{\ell}$ in \eqref{eq:hl}, the index of 
the first nonzero component of $\zh_{\ell}$ is $1 + {d}_{\ell}$. This observation can be used to find $d_{\ell}$ from an estimate of $\zh_{\ell}$.

\section{IUS parameter estimation as a sparse signal recovery problem}\label{sec:srf}
By construction, $\bA\in\mathbb{C}^{M\times \nN}$ in \eqref{MEASUREMENT_EQUATION} is a known matrix. 
On the other hand, $\zx$ and $\ze$ are unknowns and we have to obtain an estimate of $\zx$ to resolve the IUS problem.
Typically, the total number of RTs $K = \sum_{\ell = 1}^G
\tilde{N}_{\ell} \ll G$, implying $\tilde{N}_{\ell} = 0$ 
(and therefore $\zh_{\ell} = 0$) for a vast 
majority of the values $\ell \in \{1,2,\ldots,G\}$. This makes $\zx$
very sparse, motivating a sparse recovery approach for
solving the IUS problem.

Suppose we obtain a sparse estimate $\breve{\zx}$ of $\zx$ by applying a sparse recovery algorithm on \eqref{MEASUREMENT_EQUATION}. From the estimate $\breve{\zx}$, the eNodeB can extract the IUS 
information as follows. Partition $\breve{\zx}$ into $G$ sub-vectors:
\[
\breve{\zx} = [ \ \breve{\zh}_1^{\intercal} \ \ \breve{\zh}_2^{\intercal} \ \ 
\cdots \ \ \breve{\zh}_G^{\intercal} \ ]^{\intercal},
\] 
where each $\breve{\zh}_{\ell}$ is of length $N_1$. 
Then we declare $\ell \in \mathcal{L}$ only
if $\|\breve{\zh}_{\ell}\|_2 \ne 0$ and the index of the 
first nonzero component of $\breve{\zh}_{\ell}$ leads to an estimate of 
${d}_{\ell}$. 

\subsection{Performance of sparse recovery algorithm for resolving the IUS problem}
To observe the IUS parameter estimation performance by using a sparse recovery algorithm, we apply a popular approach called least absolute shrinkage and selection operator (Lasso)\cite{lasso1,lasso3}. Using the Lasso paradigm, we need to solve the following optimization problem:
\begin{align}\label{optim1}
\zx_*=\arg \min_{\zz} \ \lambda\|\zz\|_1+\frac{1}{2}\|\bA\zz-\zy\|_2^2
\end{align}
where the value of $\lambda>0$ depends on the noise level. The typical results of IUS parameter estimation by using the Lasso with similar setup in Table-\ref{tab:ex1} has been demonstrated in Table-\ref{tab:ex2}. Similar to the state of the art IUS algorithms, the performance of Lasso also depends on the code matrix. In fact, when SMUD and SRMD perform well, Lasso also performs well. On the other hand, for the selection of code matrices leading to performance deterioration of the SMUD and SRMD, Lasso also shows very clear deterioration of performance. This observation inspires us to investigate the dependency of algorithms performance on the code matrix by using the compressive sensing theory\footnote{Sparse recovery algorithms and theories are generally developed in the research area called ``Compressive sensing".}.

\subsection{The coherence parameter}
In general, the performance of sparse recovery algorithms depends on some properties of the matrix $\bA$.
Two types of metric are commonly used to characterize the properties of a matrix: i) restricted isometry property (RIP) \cite{cs6,bpn2} and (ii) mutual coherence \cite{lasso3,lasso2}. 
A matrix satisfying the restricted isometry property will approximately preserve the length of all signals up to a certain sparsity, thereby, provides performance guaranty of sparse recovery algorithms.
However, evaluating the RIP of a matrix is a computationally hard problem in general \cite{ripnp}. On the other hand, computing mutual coherence of a matrix is easy. In this respect, the coherence based results of performance guaranty of sparse recovery algorithms are appealing since they can be evaluated easily for any arbitrary matrix. In this work, we seek the performance guarantee of sparse recovery algorithms based on the mutual coherence.

The ``mutual coherence" $\mu(\bA)$ of $\bA$ is defined as \cite{lasso3}
\begin{align}\label{eq:coh}
\mu(\bA)
=\max_{i\not=j}\frac{|[\bA]_i^*[\bA]_j|}{\|[\bA]_i\|_2\|[\bA]_j\|_2},
\end{align}
where $[\bA]_i$ denotes the $i$-th column of $\bA$ and $[\bA]_i^*$ is its complex conjugate transpose.
In particular, the coherence is defined as the maximum absolute value of the cross-correlations between the normalized columns of $\bA$. When the coherence is small, the columns look very different from each other, which makes them easy to distinguish.  Thereby, it is used as a measure of the ability of sparse recovery algorithms to correctly identify the true representation of a sparse signal \cite{mu2}. The theoretical results in \cite{lasso3,lasso2} show that the performance of a sparse recovery algorithm can be improved by minimizing $\mu(\bA)$.
We validate this theoretical result for IUS application in Table-\ref{tab:ex2} where we list the values of $\mu(\bA)$ for different ZC root $u$ and $n_{cs}$. By comparing those values of $\mu(\bA)$ with the $P_s$ of Lasso, we see that the performance of Lasso increases with decreasing the value of $\mu(\bA)$.  Thereby, we can improve the performance of Lasso by minimizing the value of $\mu(\bA)$. In the following section, we show that the value of $\mu(\bA)$ can be controlled by properly designing the code matrix. 
%The Section-\ref{sec:code} will investigate how we can control 
%$\mu(\bA)$ to our advantage. 

\begin{table*}
\centering \caption{IUS user detection probabilities by Lasso and mutual coherence of $\bA$ for different values of of $u$ and $n_{cs}$. SNR=$10$ $\zd$B, total IUS users $K=3$ and $P_s$ denotes the probability of successfully detecting the users.}
\begin{tabular} {|c| c | c | c | c |c|c|c|c|} 
\hline
& \multicolumn{5}{|c |}{ZC root $u=1$}&\multicolumn{3}{c |}{ZC root  $u=2$}\\
\cline{2-9}
 & $n_{cs}=11$ & $13$ & $15$&$17$&$19$&$n_{cs}=11$ & $13$ &$17$\\
\hline\hline
$P_s$& $0.71$ & $0.7$&$0.99$&$0.99$&$0.99$&$0.27$&$0.68$&$0.71$\\
% 2 & $0.27$ & $0.67$&$0.92$&$0.92$&$0.93$\\
\hline
$\mu(\bA)$&  $0.994$ & $0.998$ & $0.969$&$0.969$&$0.969$&$1.0$ & $0.998$ &$0.993$\\
%\hline
%\multicolumn{8}{|c |}{Mutual Coherence $\mu(\bA)$}\\
% \hline
% \multicolumn{5}{|c |}{ZC root  $u=1$}&\multicolumn{3}{c |}{ZC root  $u=2$}\\
%\cline{1-8}
%  $n_{cs}=11$ & $13$ & $15$&$17$&$19$&$n_{cs}=11$ & $13$ &$17$\\
%\hline\hline
%  $0.994$ & $0.998$ & $0.969$&$0.969$&$0.969$&$1.0$ & $0.998$ &$0.993$\\
%
%% \parbox{1.1cm}{ ZC root ($u$)} &\multicolumn{5}{c |}{Mutual coherence $\mu(\bA)$}\\
%%\cline{2-6}
%% &$n_{cs}=11$ & $13$ & $15$&$17$&$19$\\
%%\hline\hline
%%1 & $0.994$ & $0.998$&$0.969$&$0.969$&$0.969$\\
%% 2& $1.0$ & $0.998$&$0.9968$&$0.9939$&$1.0$\\
\hline
\end{tabular}\label{tab:ex2}
\end{table*}

%\begin{table}
%\centering \caption{RA user detection probabilities by Lasso and mutual coherence of $\bA$ for different values of of $u$ and $n_{cs}$. SNR=$10$ $\zd$B, total RA users $K=3$ and $P_s$ denotes the probability of successfully detecting the users.}
%\begin{tabular} {| c | c | c | c |c|c|} 
%\hline
% \parbox{1.1cm}{  ZC root ($u$)} &\multicolumn{5}{c |}{$P_s$ of Lasso}\\
%\cline{2-6}
%  &$n_{cs}=11$ & $13$ & $15$&$17$&$19$\\
%\hline\hline
%1 & $0.71$ & $0.7$&$0.99$&$0.99$&$0.99$\\
% 2 & $0.27$ & $0.67$&$0.92$&$0.92$&$0.93$\\
%\hline
%\hline
% \parbox{1.1cm}{ ZC root ($u$)} &\multicolumn{5}{c |}{Mutual coherence $\mu(\bA)$}\\
%\cline{2-6}
% &$n_{cs}=11$ & $13$ & $15$&$17$&$19$\\
%\hline\hline
%1 & $0.994$ & $0.998$&$0.969$&$0.969$&$0.969$\\
% 2& $1.0$ & $0.998$&$0.9968$&$0.9939$&$1.0$\\
%\hline
%\end{tabular}\label{tab:ex2}
%\end{table}

% i.e.,
%\begin{align}\label{eq:js}
%\mI_s:=\{\ell\in\{1,2,\cdots G\}:u_{\ell}=s\},  \ \forall s\in \mathcal{U}.
%\end{align}

%We first derive an expression for $\mu(\bA)$ by assuming that all codes in the code matrix $\bC$ are generated from same root, i.e., $u_{\ell}=u_{\ell+1};\forall \ell\in\{1,2, \cdots G-1\}$ in \eqref{eq:code}. Second, we derive a relation for $\mu(\bA)$ by assuming that every code of  $\bC$ is generated from different root, i.e., $u_{\ell}\not=u_{m};\forall \ell\not=m$. Finally, we combine both results to provide a bound of $\mu(\bA)$ for general $\bA$.
\section{Single root code matrix design}
In this section, we derive an expression for $\mu(\bA)$ by assuming that all codes in the code matrix $\bC$ are generated from the same ZC root, i.e., we assume that $u_{\ell}=\tilde{u};\forall \ell\in\{1,2, \cdots G\}$ in \eqref{eq:code}. Consequently, we propose a code matrix design procedure that can ensure the optimum value of $\mu(\bA)$. In the next section, we extend the code matrix design procedure for multiple roots.

\subsection{Coherence of $\bA$ for single root}

 The following lemma gives the coherence property of $\bA$.

\begin{Lem}\label{th2} Assume that $u_{\ell}=\tilde{u};\forall \ell\in\{1,2, \cdots G\}$ in \eqref{eq:code}. Furthermore, $\pi\sqrt{2}\le M\le  N/2$. Then it holds that
\begin{align}\label{eq:mu}
\mu(\bA)=
\left |
\frac{\mathrm{sinc} (g^{(\tilde{u})} M/N) }{
\mathrm{sinc} (g^{(\tilde{u})}/N)}
\right|,
\end{align}
 where $\mathrm{sinc}(x)=\sin(\pi x)/(\pi x)$ and
$g^{(\tilde{u})}=\min\{1,\zeta(\tilde{u})\}$.
Here we define
\begin{align}
\label{eq:beta}
\zeta(\tilde{u}) &:=\min_{\substack{\ell\not=m\\ 1\le p,k\le N_1}}\left\{\left(\frac{n_{cs} \ \tilde{u} N}{M}(\ell-m)+(p-k)\right) \ \mathrm{mod} \  N\right\}
\end{align}
with $m,\ell\in\{1,2,\cdots G\}$.
%where as usual, $(k \ \mathrm{mod} \ N) := k - N \cdot \lfloor k/N 
%\rfloor $, with $\lfloor r \rfloor$ denoting the largest
%integer less than or equal to $r$. 
\end{Lem}
{\bf Proof:} See Appendix-\ref{app1}. \feop

In a typical LTE system \cite{lte2} the values of $M=839$ and $N=6144$. Therefore, the assumption of Lemma-\ref{th2} i.e., $\pi\sqrt{2}\le M\le  N/2$ holds in practice. 
Under the assumptions of Lemma-\ref{th2}, the value of $\mu(\bA)$ is the smallest when $g^{(\tilde{u})}=1$. If the value of $\zeta(\tilde{u})$ becomes smaller than one then it will increase the value of matrix coherence. The code design procedure proposed here will aim to maintain $\zeta(\tilde{u})\ge1$.
%Since $\zb_{\ell}$ depends on the values of $u$ and $c_{ns}$, thereby, we can control the value of coherence by properly choosing $u$ and $c_{ns}$. 

In this following sections, we first demonstrate the conventional procedure of code matrix design. We show that the procedure may not achieve the minimum value of coherence of $\bA$. Finally, we demonstrate an efficient code matrix design procedure.

\subsection{Conventional procedure of generating RA code matrix (CRA)}\label{tab:ra1}
 The conventional procedure of generating RA code sequences has been described in \cite{lte4,lte9}. For the given specification of an LTE system, we need to compute the smallest integer for $n_{cs}$ that satisfies \eqref{eq:ncs}. For a given specification of the LTE system, we take $\hat{n}_{cs}$ as the smallest integer satisfying \eqref{eq:ncs}. Set $n_{cs}=\hat{n}_{cs}$ and compute $G=\lfloor M/n_{cs}\rfloor$ which is the maximum number of codes that can be generated from a single root. We then choose an arbitrary non-zero positive integer for the root $\tilde{u}$ such that $\tilde{u}<M$ \cite{lte4}. Subsequently, we generate $G$ number of RA codes $\{\zc_{\ell}\}_{\ell=1}^G$ by using \eqref{eq:zc2}. However, this procedure may produce a code matrix $\bA$ with bad value of $\mu(\bA)$. Such examples are readily constructed. Consider the LTE system with $N=6144, N_g=768$ and $M=839$ and OFDM symbol sampling interval is $130$ ns. The cell radius $\gamma=1.5$ km and $N_1=105$. Using \eqref{eq:ncs}, we obtain a lower bound $\hat{n}_{cs}=13$. Suppose we want to generate total $50$ codes. By using those values, we construct a code matrix $\bC$ by applying the above procedure where we set $\tilde{u}=1$. We found that $ \zeta(1)=0.199$ and consequently $\mu(\bA)=0.9988$.

%\begin{table}[!t]
%\renewcommand{\arraystretch}{1.3}
%\centering \caption{Conventional procedure of RA code generation (CRA)}\label{tab:ra1}
%\begin{tabular}{l}
% \hline
%{\bf Input:} Values of a ZC root $s$, and $\hat{n}_{cs}$.\\
%{\bf Initialization:} Set $\bC$ is empty, $n_{cs}=\hat{n}_{cs}$, $G=\lfloor M/n_{cs}\rfloor$.\\
%\ \ \ 1. {\bf For} $\ell=1:G$\\
%\ \ \ \ \ 2. Generate $\zc_{\ell}^{(s)}$ using \eqref{eq:zc2}.\\
%\ \ \ \ \ 3. Update $\bC=[\bC \ \ \zc_{\ell}^{(s)}]$.\\
%\ \ \ 4. End for.\\
%{\bf 5. Output:} \ Code matrix $\bC$.\\
%\hline
%\end{tabular}
%\end{table}

\subsection{Coherence based code generation (CCG)}\label{sec:ccg}
 
%In general, the sparse recover algorithms can be classified into two major categories: optimization-based methods and greedy methods. It is well known \cite{omp1,mud6} that the performance of both techniques depend on a property of the matrix $\bA$ called ``coherence"; lower matrix coherence ensures better sparse signal recovery guarantee.

%\subsubsection{Code generation}

%The code generation procedure described here will aim to minimize the coherence of $\bA$.
Recall that the minimum value of $\mu(\bA)$  in \eqref{eq:mu}  %will attain at $g^{(\tilde{u})}=1$ which
 can be obtained by making $\zeta(\tilde{u})\ge 1$, i.e., 
for every $p,k\in\{1,2,\cdots N_1\}$ and $\ell,m\in\{1,2,\cdots G\}$ with $\ell\not=m$, we have to satisfy
\begin{align}\label{eq:ncon1}
\left\{\left(\frac{n_{cs} \ \tilde{u} N}{M}(\ell-m)+(p-k)\right) \ \mathrm{mod} \  N\right\}\ge 1.
\end{align}
The following proposition gives a sufficient condition to fulfill the requirement in \eqref{eq:ncon1}.

\begin{Pro}
%The sufficient condition to satisfy  \eqref{eq:ncon1} for every $p,k\in\{1,2,\cdots N_1\}$ and $\ell,m\in\{1,2,\cdots G\}$ with $\ell\not=m$ is given by
If
\begin{align}
\label{eq:con23} N_1\le \frac{n_{cs}\tilde{u}N}{M}\le \frac{N-N_1}{G-1},
\end{align}
 then  \eqref{eq:ncon1} holds for every $p,k\in\{1,2,\cdots N_1\}$ and $\ell,m\in\{1,2,\cdots G\}$ with $\ell\not=m$.
\end{Pro} 
{\bf Proof:} It is sufficient to show that if \eqref{eq:con23} holds then
\begin{align}\label{eq:ncon4}
1\le \left|\frac{n_{cs} \ \tilde{u} N}{M}(\ell-m)+(p-k)\right|\le N-1
%\left|\frac{n_{cs} \ \tilde{u} N}{M}(\ell-m)+(p-k)\right| \in[1,N-1]
\end{align}
 for any $p,k\in\{1,2,\cdots N_1\}$ and $\ell,m\in\{1,2,\cdots G\}$ with $\ell\not=m$.
% , i.e,
%\begin{align}\label{eq:ncon4}
%1\le \left|\frac{n_{cs} \ \tilde{u} N}{M}(\ell-m)+(p-k)\right|\le N-1.
%\end{align}

Suppose that \eqref{eq:con23} holds. Then using reverse triangle inequality \cite{rte}, we see that
\begin{align}\label{eq:ncon5}
&\left|\frac{n_{cs} \ \tilde{u} N}{M}(\ell-m)+(p-k)\right|\nonumber\\
&\ge \left|\frac{n_{cs} \ \tilde{u} N}{M}(\ell-m)\right|-|(p-k)|\nonumber\\
&\ge \frac{n_{cs} \ \tilde{u} N}{M}-(N_1-1)\nonumber\\
&\ge 1
\end{align}
where the last inequality follows from the first inequality of \eqref{eq:con23}. 

On the other hand, triangle inequality 
implies
\begin{align}\label{eq:ncon6}
&\left|\frac{n_{cs} \ \tilde{u} N}{M}(\ell-m)+(p-k)\right|\nonumber\\
&\le \left|\frac{n_{cs} \ \tilde{u} N}{M}(\ell-m)\right|+|(p-k)|\nonumber\\
&\le \frac{n_{cs} \ \tilde{u} N (G-1)}{M}+(N_1-1)\nonumber\\
&\le N-N_1-N_1+1= N-1
\end{align}
where we use the second inequality of \eqref{eq:con23}. Combining \eqref{eq:ncon5} and \eqref{eq:ncon6} we get \eqref{eq:ncon4}.
\feop

{\bf {\em Remark 1:}} Note that for some given values of $n_{cs}$ and $\tilde{u}$, eq. \eqref{eq:con23} gives an upper bound on the maximum number of codes that can be generated i.e., 
\begin{align}\label{eq:con2}
G\le 1+\frac{M(N-N_1)}{N {n}_{cs}\tilde{u}}.
\end{align}

Table-\ref{tab:ra2} shows the proposed CCG algorithm. Here $\hat{n}_{cs}$ denotes the value of lower bound of $n_{cs}$ that satisfies \eqref{eq:ncs}. Unlike the CRA algorithm (Section-\ref{tab:ra1}), we do not use the value of $\hat{n}_{cs}$ directly for code generation. Instead, we use $\hat{n}_{cs}$ to choose an appropriate value of ${n}_{cs}$ in Step-1 which satisfies the lower bound in \eqref{eq:con23}. Given ${n}_{cs}$, we use \eqref{eq:con2} to compute the value of $G$ in Step-2. Subsequently, the algorithm generates $G$ number of codes using Step-3 to Step-6.

\begin{table}[!t]
\renewcommand{\arraystretch}{1.3}
\centering \caption{Coherence based code generation (CCG) for single ZC root}\label{tab:ra2}
\begin{tabular}{l}
 \hline
{\bf Input:} The value of $\hat{n}_{cs}$ and a ZC root $\tilde{u}$.\\
{\bf Initialization:} Set $\bC$ is empty.\\
\ \ \ 1. Find the smallest positive integer ${n}_{cs}$ such that \\
\ \ \ \ \ \  $\hat{n}_{cs}\le n_{cs}$ and satisfies lower bound of \eqref{eq:con23} i.e., $\frac{MN_1}{N}\le  {n}_{cs} \tilde{u}$.\\
\ \ \ 2. Compute $G$ using \eqref{eq:con2}: $G= \lfloor 1+\frac{M(N-N_1)}{N {n}_{cs}\tilde{u}}\rfloor.$\\
\ \ \ 3. {\bf For} $\ell=1:G$\\
\ \ \ \ \ 4. Generate $\zc^{(\tilde{u})}_{\ell}$ using \eqref{eq:zc2}.\\
\ \ \ \ \ 5. Update $\bC=[\bC \ \ \zc^{(s)}_{\ell}]$.\\
\ \ \ 6. End for.\\
{\bf 7. Output:} \ Code matrix $\bC$.\\
\hline
\end{tabular}
\end{table}

\section{Multiple root code matrix design} %Coherence of $\bA$ for different roots}
In practice, the code matrix $\bC$ in \eqref{eq:code} can be generated by using multiple ZC roots. We denote $\mathcal{U}$ as the set of all roots that has been used to generate $\bC$ i.e.  $u_{\ell}\in \mathcal{U};\forall \ell\in\{1,2,\cdots G\}$. Let $\mI_u$ be the set of all column indices of $\bC$ that are generated from the root $u$. To give an example, assume that $\bC$ has been generated from the two roots $\{1,2\}$, where the first $3$ columns of $\bC$ are generated using root number $1$. Hence, $\mathcal{U}:=\{1,2\}$ and $\mI_1:=\{1,2,3\}$. Hence, in \eqref{eq:code} we have $u_{\ell}=1;\ell=1,2,3$  and $u_{\ell}=2;\ell=4,\cdots G$. Thus, in this case 
$$\bC = [ \ \zc^{(1)}_1 \ \ \zc^{(1)}_2 \ \ \zc^{(1)}_3 \ \ \zc^{(2)}_4 \ \cdots \ \ \zc^{(2)}_G \ ].$$
Let $\bB_u$ be the matrix which is constructed by concatenating the matrices $\tilde{\bE}_{\ell}; \ell\in\mI_u$. For example, if the first three columns of $\bC$ are generated from the root $u$ then $\mI_u:=\{1,2,3\}$ and $\bB_u=[\tilde{\bE}_{1} \ \ \tilde{\bE}_{2} \ \ \tilde{\bE}_{3}]$. In this way, we can view $\bA$ as a concatenation of matrices $\bB_u;u\in \mathcal{U}$. 

To derive an expression for $\mu(\bA)$, we require the following definition. For any two matrices $\bU\not=\bD$, we extend the definition of mutual coherence in \eqref{eq:coh} to ``block coherence" as
\begin{align}\label{eq:coh2}
\hat{\mu}(\bU,\bD)=\max_{j,k}\frac{|[\bU]_j^*[\bD]_k|}{\|[\bU]_j\|_2 \ \|[\bD]_k\|_2}.
\end{align}
where $[\bD]_k$ denotes the $k$-th column of $\bD$. 
Therefore
\begin{align}\label{eq:call3}
\mu(\bA)=\max\{\max_{p\in\mathcal{U}}\mu({\bB}_p),\max_{u1\not=u2; \  u1,u2\in\mathcal{U} }\hat{\mu}({\bB}_{u1},{\bB}_{u2})\}.
\end{align}
In the following, we develop a code design procedure so that  we can maintain $\mu(\bA)$ to its minimum value i.e.,
\begin{align}\label{eq:gmu}
\mu(\bA)=\left |
\frac{\mathrm{sinc} (M/N) }{
\mathrm{sinc} (1/N)}
\right|.
\end{align}
We assume that the following condition holds for any $p\in\mathcal{U}$
\begin{align}\label{eq:alls}
\mu(\bB_p)=
\left |
\frac{\mathrm{sinc} (M/N) }{
\mathrm{sinc} (1/N)}
\right|.
\end{align}
In particular, we can satisfy \eqref{eq:alls} by constructing $\bB_p$ using the CCG method in Section-\ref{sec:ccg}. Next, we see that
\begin{align}\label{eq:be}
\max_{u1\not=u2; \  u1,u2\in\mathcal{U} }\hat{\mu}({\bB}_{u1},{\bB}_{u2})=\max_{\substack{\ell \in\mI_{u1}; \ m\in\mI_{u2} \\ u1\not=u2; \  u1,u2\in\mathcal{U} }}\hat{\mu}(\tilde{\bE}_{\ell},\tilde{\bE}_{m}).
\end{align}
It is difficult to develop an exact expression for \eqref{eq:be}. Nevertheless, we provide an upper bound for the value of \eqref{eq:be}.  
By using the definition of block coherence in \eqref{eq:coh2}, we get 
\begin{align}\label{eq:coh4}
\hat{\mu}(\tilde{\bE}_{\ell},\tilde{\bE}_{m})=\max_{1\le k,p\le N_1}\frac{1}{M}|[\tilde{\bE}_{\ell}]_k^*[\tilde{\bE}_{m}]_p|.
\end{align}
 Next we give an upper bound on the right hand side of \eqref{eq:coh4}.
\begin{Pro}\label{pro1} For any $k,p\in\{1,2,\cdots N_1\}$ with $\ell\in\mI_{u1}; \ m\in\mI_{u2}$; $ u1,u2\in\mathcal{U} $ and $u1\not=u2$, it holds that
\begin{align}
&\frac{1}{M}|[\tilde{\bE}_{\ell}]_k^*[\tilde{\bE}_{m}]_p|\nonumber\\
\label{eq:elm}&= \frac{1}{M}\bigg |\sum_{n=0}^{M-1}{\exp\Big\{-\irm \frac{\pi}{M}(u2-u1)}\nonumber\\
& \ \ \ \ \ \ \ \ \ \ \ \ \ \ \ \ \  \ \ \ \ \ \ \ \ \ \left\{n+\vartheta(u1,u2,\ell,m,p,k)\right\}^2 \Big\}\bigg |\\
\label{eq:app12} &\mathrm{where,} \ \vartheta(u1,u2,\ell,m,p,k)\nonumber\\
&  =\frac{1}{2}+\frac{1}{(u2-u1)}\bigg\{n_{cs}\Big(u2(m-1)\nonumber\\
& \ \ \ \ \ \ \ \ \ \ \ \ \ \ \ \ \  \ \ \ \ \ \ \  \ \ \ -u1(\ell-1)\Big)+(p-k)\frac{M}{N}\bigg\}.
\end{align}
\end{Pro}
{\bf Proof:} See Appendix-\ref{app3}. \feop

\begin{Lem} \label{fzc}
Suppose the following conditions hold for some $k,p\in\{1,2,\cdots N_1\}$ with $\ell\in\mI_{u1}; \ m\in\mI_{u2}$ and $u1\not=u2$:
\begin{itemize}
\item[C1.] $M$ is an odd prime integer.
\item[C2.] $(u2-u1)$ is an even integer.
\item[C3.] $\vartheta(u1,u2,\ell,m,p,k)$ is integer valued.
\end{itemize}
 Then the following relation holds
\begin{align}\label{eq:e12}
\frac{1}{M}|[\tilde{\bE}_{\ell}]_k^*[\tilde{\bE}_{m}]_p|= \frac{1}{\sqrt{M}}.
\end{align}
\end{Lem}
{\bf Proof:} Under conditions C1-C3, the sequence
\begin{align}\label{eq:l1}
\zz(n)=\exp\left(-\irm \frac{\pi}{M}(u2-u1)\left\{n+\vartheta(u1,u2,\ell,m,p,k)\right\}^2 \right)
\end{align}
with $n=0,1,2,\cdots M-1$ satisfies $|\sum_{n=0}^{M-1}\zz(n)|=\sqrt{M}$ \cite{zc3}. Thereby, \eqref{eq:e12} holds.\feop

{\bf {\em Remark 2:}} Under conditions C1-C3 of Lemma-\ref{fzc}, the sequence $\zz(n)$ (defined in \eqref{eq:e12}) is
known as the Frank-Zadoff-Chu (FZC) sequence \cite{zc1,zc10,zc3}. 

Note that in LTE system $M=839$. Thus, the first condition of the Lemma-\ref{fzc} holds. Furthermore, we can fulfill the second condition by appropriately choosing the ZC root sequences. However, the third condition may not hold for all $k,p\in\{1,2,\cdots N_1\}$. We observe that $|\sum_{n=0}^{M-1}\zz(n)|\approx\sqrt{M}$, whenever $\vartheta(u1,u2,\ell,m,p,k)$ is approximately an integer. In practice, the value of \eqref{eq:elm} remains close to $1/\sqrt{M}$ for any feasible value of $\vartheta(u1,u2,\ell,m,p,k)$ defined in \eqref{eq:app12}. Furthermore, for a typical LTE system ($M=839, N=6144$)
$$
\frac{1}{\sqrt{M}}\ll\left |
\frac{\mathrm{sinc} (M/N) }{
\mathrm{sinc} (1/N)}
\right|.$$
Hence,
\begin{align}\label{eq:be2}
\max_{\substack{\ell\in\mI_{u1}; \ m\in\mI_{u2} \\ u1\not=u2; \  u1,u2\in\mathcal{U} }}\hat{\mu}(\tilde{\bE}_{\ell},\tilde{\bE}_{m})\le \left |
\frac{\mathrm{sinc} (M/N) }{
\mathrm{sinc} (1/N)}
\right|
\end{align}
holds with high probability, and thus we should be able to find roots $u1$ and $u2$ satisfying \eqref{eq:gmu}. Nevertheless, to be more precise, we develop a systematic procedure to choose the ZC roots so that \eqref{eq:gmu} holds. 

The root selection problem can be described formally in the following way. Suppose, total $t$ number of columns of $\bC$ have already been generated from the set of roots $\mathcal{U}$ such that \eqref{eq:gmu} is satisfied. Given $\mathcal{U}$ and $\mI_{u1};u1\in \mathcal{U}$ we need to choose another root $u2$ and generate additional columns of $\bC$ so that \eqref{eq:gmu} holds.

Note that for some given values of $u2$ and $n_{cs}$ satisfying the lower bound of \eqref{eq:con23}, we can generate maximum $G_{u2}=\lfloor \frac{M(N-N_1)}{N \ n_{cs}u2}+1 \rfloor$ number of columns of $\bC$ (see \eqref{eq:con2}). Denote $\mI_{u2}=\{t+1,t+2\cdots t+G_{u2}]$. Since the columns are generated by satisfying \eqref{eq:con23}, we have
$$
\mu(\bB_r)=
\left |
\frac{\mathrm{sinc} (M/N) }{
\mathrm{sinc} (1/N)}
\right|.
$$
Then according to Proposition-\ref{pro1}, it is sufficient to check
\begin{align}\label{eq:be3}
&\max_{\substack{p,k\in\{1,2,\cdots N_1\}\\ m\in\mI_{u2}, \  \ell\in\mI_{u1}; \ \forall u1\in\mathcal{U} }}\frac{1}{M}\bigg |\sum_{n=0}^{M-1}{\exp\Big(-\irm \frac{\pi}{M}(u2-u1)}\nonumber\\
& \ \ \ \ \ \ \ \left\{n+\vartheta(u1,u2,\ell,m,p,k)\right\}^2 \Big)\bigg |
\le \left |
\frac{\mathrm{sinc} (M/N) }{
\mathrm{sinc} (1/N)}
\right|.
\end{align}
Based on the idea, the multiple root code matrix generation procedure has been described in Table-\ref{tab:ra3}. Suppose that we want to generate total $G$ number of codes. The algorithm initiates with the ZC root $u=1$. Given the value of $u$, and $\hat{n}_{cs}$ as the lower bound of $n_{cs}$, we compute the values of ${n}_{cs}$ in Step 1 and $G_u$ in Step-2. In Step-4, we check that whether the condition in  \eqref{eq:be3} is satisfied. If the condition is satisfied then we concatenate $G_u$ number of codes for $\bC$ in Step-5 to Step-10. In Step-11, we add the active root $u$ with the set $\mathcal{U}$.

\begin{table}[!t]
\renewcommand{\arraystretch}{1.3}
\centering \caption{Coherence based code generation (CCG) for multiple ZC root}\label{tab:ra3}
\begin{tabular}{l}
 \hline
{\bf Input:} The value of $G$ and the value of $\hat{n}_{cs}$.\\
{\bf Initialization:} Set $\bC$ is empty, $\mathcal{U}$ is empty, $u=1$ and $t=0$.\\
{\bf repeat}\\
\ \ \ 1. Find the smallest positive integer ${n}_{cs}$ such that \\
\ \ \ \ \ \  $\hat{n}_{cs}\le n_{cs}$ which satisfies lower bound of \eqref{eq:con23} i.e., $\frac{MN_1}{N}\le  {n}_{cs} u$.\\
\ \ \ 2. Compute $G_u$ using \eqref{eq:con2}: $G_u= \lfloor 1+\frac{M(N-N_1)}{N{n}_{cs}u}\rfloor.$\\
\ \ \ 3. Set $\mI_u:=\{t+1,t+2\cdots t+G_u\}$.\\
\ \ \ 4. If the condition in \eqref{eq:be3} does not\\
\ \ \ \ \ satisfy then goto Step 12.\\
\ \ \ 5. {\bf For} $\ell=1:G_u$\\
\ \ \ \ \ 6. Generate $\zc^{(u)}_{\ell}$ using \eqref{eq:zc2}.\\
\ \ \ \ \ 7. Update $\bC=[\bC \ \ \zc^{(u)}_{\ell}]$.\\
\ \ \ \ \ 8. $t=t+1$.\\
\ \ \ \ \ 9. If $t=G$ then goto Step 13.\\
\ \ \ 10. End for.\\
\ \ \ 11. Set $\mathcal{U}:=\{\mathcal{U}, \ u\}$.\\
\ \ \ 12. $u=u+1$.\\
{\bf Continue to Step 1.}\\
{\bf 13. Output:} \ Code matrix $\bC$.\\
\hline
\end{tabular}
\end{table}

\begin{figure}[]
\centering
\subfloat[SMUD \cite{rang1}]{\includegraphics[width=8.5cm]{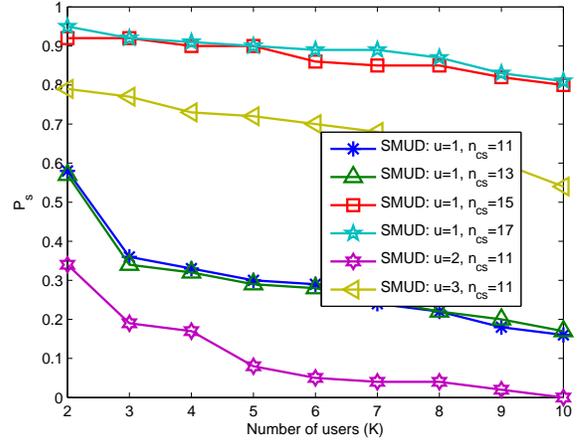}}\\
\subfloat[SRMD \cite{rang5}]{\includegraphics[width=8.5cm]{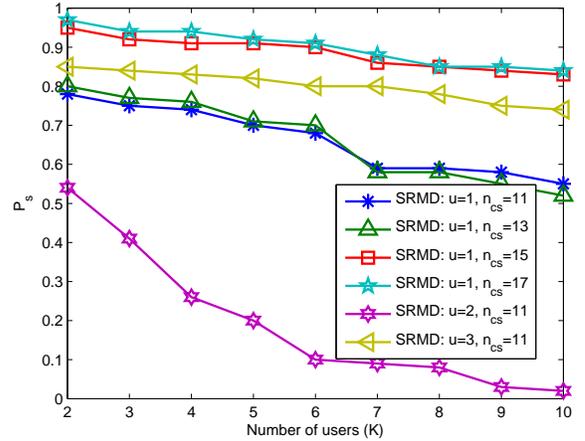}}\\
%\subfloat[RA-GLRT \cite{lte2}]{\includegraphics[width=8.5cm]{figure/fig113.eps}}
\subfloat[Lasso]{\includegraphics[width=8.5cm]{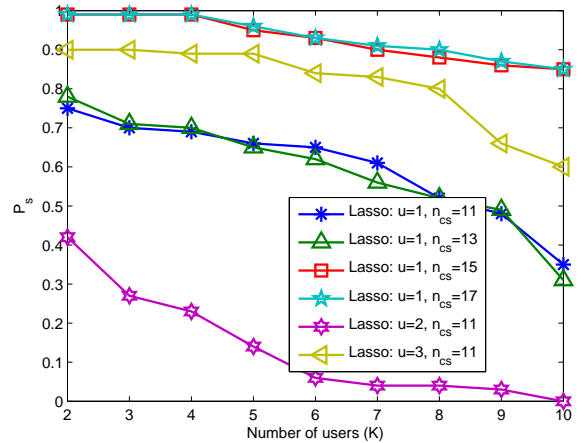}}
\caption{ Code detection performance as a function of number of users for different values of ZC root $u$ and $n_{cs}$. Signal SNR=$10$ dB. The CRA matrices are generated with $n_{cs}\in\{11,13\}$ and $u\in\{1,2,3\}$. The CCG matrices are generated using $n_{cs}\in\{15,17\}$ and $u=1$.} \label{fig1}
\end{figure}

\section{Simulation Results}\label{sec:exp}

We simulate an LTE system similar to \cite{lte2} where $N=6144$,
and cyclic prefix length $N_g = 768$. 
The subcarrier frequency spacing is $1.25$ kHz, and the sampling interval 
$T_s=130$ ns. The PRACH consist of $M=839$ 
adjacent subcarriers. The wireless cell radius is 
$1.3$ km, which corresponds to $D=70$. The wireless channels are
modeled according to a mixed channel 
model specified by the ITU IMT-2000 standards: Ped-A, Ped-B, and 
Veh-A. For each RT, the simulator selects one of the above channel 
models uniformly at random.
The mobile speed varies in the interval $[0,5]$ m/s for Ped-A, Ped-B 
channels, and $[5,20]$ m/s for Veh-A. The channel impulse response of 
the RTs have a maximum order of $P_{\max}=35$ taps \cite{lte2}.
 Similar to \cite{rang1,rang3}, we assume that eNodeB 
has an approximate knowledge about $P_{\max}$, and we set 
$N_1=P_{\max}+D$ in \eqref{ONE_TERMINAL_RELATION}. 
The RA codes are the Zadoff-Chu (ZC) sequences of length $839$. The lower bound of $n_{cs}$ for the ZC sequence is calculated by using \eqref{eq:ncs}, and we get $n_{cs}\ge 11$. The number 
$G$ of available RA codes in the matrix $\bC$ for contention based random access is $50$.
Recall that, at a particular random access opportunity, the set of active RA code indices is $\mathcal{L}$.
Let $\hat{\mathcal{L}}$ be the set of code indices detected by an 
algorithm. The probability that
${\mathcal{L}}=\hat{\mathcal{L}}$, denoted by $P_s$, is used 
to quantify the merit of the algorithm \cite{mud4}. 
The  signal to noise ratio (SNR) is defined as 
$\mathrm{SNR}=10\log_{10} ( \sigma_h^2 / \sigma_e^2 )$, 
where $\sigma_h^2$ is the variance of a channel tap, and 
$\sigma_e^2$ is the variance of a component of $\ze$, 
respectively \cite{rang3}. For the Lasso algorithm in \eqref{optim1}, we set $\lambda=\sqrt{8\sigma_e^2(1+\alpha)\ln(G.N_1)}$ \cite{lasso2}, where $\alpha=4$.
The following results are based on $200$ independent 
Monte-Carlo simulations.

\begin{figure}[htb]
\centering
\includegraphics[width=8.5cm]{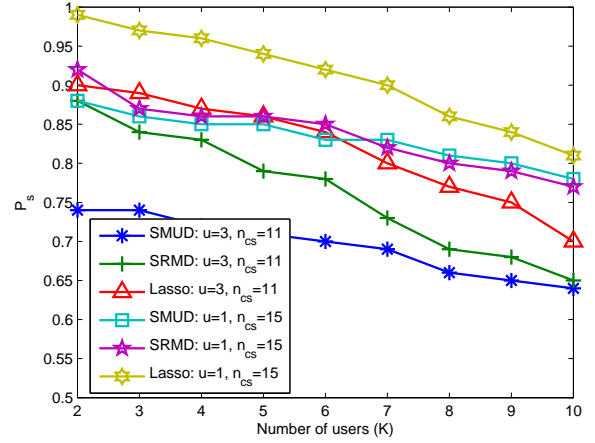}%{figure/42.eps}
\caption{Code detection performance by different algorithms at SNR=$5$ dB. The CRA matrix is generated with $n_{cs}=11$ and $u=3$. The CCG matrix is generated using $n_{cs}=15$ and $u=1$. } 
\label{fig2}
\end{figure}

In Figure-\ref{fig1}, we compare the code detection performance of different algorithms with different code matrices generated by the CRA and CCG methods. For the LTE configuration under consideration, the conventional code matrix generation procedure i.e., CRA method (see Section-\ref{tab:ra1}) suggests using $n_{cs}=11$ and we can set the value of ZC root arbitrarily. Hence, we choose three different values of ZC root i.e., $u\in\{1,2,3\}$ to illustrate performance of the IUS algorithms. In contrast, the CCG method, described in Section-\ref{sec:ccg}, suggests using $u=1$ and $n_{cs}\ge15$. Thus, we use two different values of $n_{cs}\in\{15,17\}$. Note that we apply the multiple root CCG method (Table-\ref{tab:ra3}) for code generation whenever $n_{cs}\ge17$. As can be seen in Figure-\ref{fig1}, the code detection probability of any particular algorithm remains almost similar for any value of $n_{cs}\in\{15,17\}$ with $u=1$. However, a larger value of $n_{cs}$ decreases the value of $G_u$ (see \eqref{eq:con2}), that is the maximum number of codes that can be generated from the single root $u$ with the CCG algorithm. Hence, we shall use $n_{cs}=15$ for CCG algorithm in the following experiments. As can be seen, all algorithms exhibit their optimum performances with the CCG code matrices. For example, the code detection probabilities of SMUD algorithm in Figure-\ref{fig1}(a) are $0.92$ and $0.8$ with the CCG matrix ($n_{cs}=15$) for $2$ and $10$ users respectively. In contrast, SMUD exhibits different types of performance for three different values of $u$ with the CRA matrices. The code detection probabilities remain $0.6$, $0.35$ and $0.78$ with $2$ users for $n_{cs}=11$ and $u=1,2$ and $3$ respectively. As expected the performance deteriorates with the increase in the number of users. With $10$ users, the $P_s$ are $0.17, 0.01$ and $0.54$ receptively for $n_{cs}=11$ and $u=1,2$ and $3$. The SRMD algorithm in Figure-\ref{fig1}(b) also shows similar performance with the code matrices. We have stated in Section-\ref{sec:srf} that the code detection problem can be solved by using a sparse recovery algorithm. The results in Figure-\ref{fig1}(c) justify our claim. Similar to the state of the art IUS algorithms, the Lasso can detect RA codes efficiently. To illustrate the robustness of the IUS algorithms with the CCG matrix in noisy environment, we present Figure-\ref{fig2}, where we evaluate the code detection performance of IUS algorithms in relatively lower SNR. As can be seen, the algorithms still perform better with the CCG code matrix. 

\begin{figure}[htb]
\centering
\includegraphics[width=8.5cm]{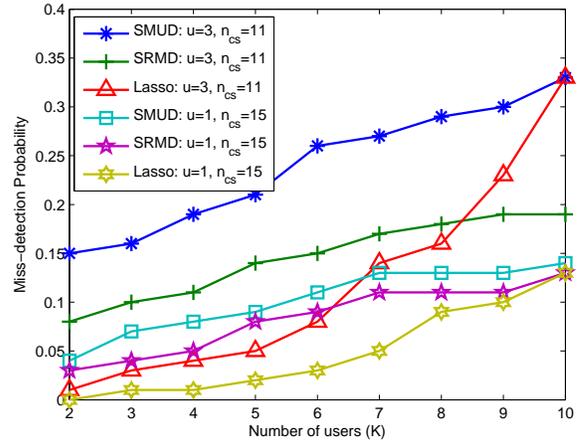}%{figure/42.eps}
\caption{Miss-detection probability by different IUS algorithms at SNR=$10$ dB. The CRA matrix is generated with $n_{cs}=11$ and $u=3$. The CCG matrix is generated using $n_{cs}=15$ and $u=1$. } 
\label{fig4}
\end{figure}

We examine the miss-detection 
probability $P_{md}$ of different IUS algorithms for different code matrices in Figure-\ref{fig4}. Here `miss-detection' occurs when 
an algorithm includes the index of an inactive code into
$\hat{\mathcal{L}}$. As can be seen the miss-detection probability of all IUS algorithms are higher with the CRA matrix compared to the CCG matrix. For example, with $6$ users the miss-detection probability of Lasso are  $0.08$ and $0.03$ for CRA and CCG code matrices respectively. The miss-detection probability increases with increasing the number of active users. With $9$ users, the miss-detection probability of Lasso are  $0.23$ and $0.1$ respectively for CRA and CCG code matrices. Both SMUD and SRMD algorithms also perform similarly. 

%
%\begin{figure}[htb]
%\centering
%\includegraphics[width=8.5cm]{figure/fig13.eps}%{figure/42.eps}
%\caption{MSE of power estimate by different algorithm with different code matrices. SNR=$10$ dB. The CRA matrix is generated with $n_{cs}=11$ and $u=3$. The CCG matrix is generated using $n_{cs}=15$ and $u=1$. } 
%\label{fig3}
%\end{figure}

\begin{figure*}[]
\centerline{\subfloat[ ]{\includegraphics[width=8.5cm]{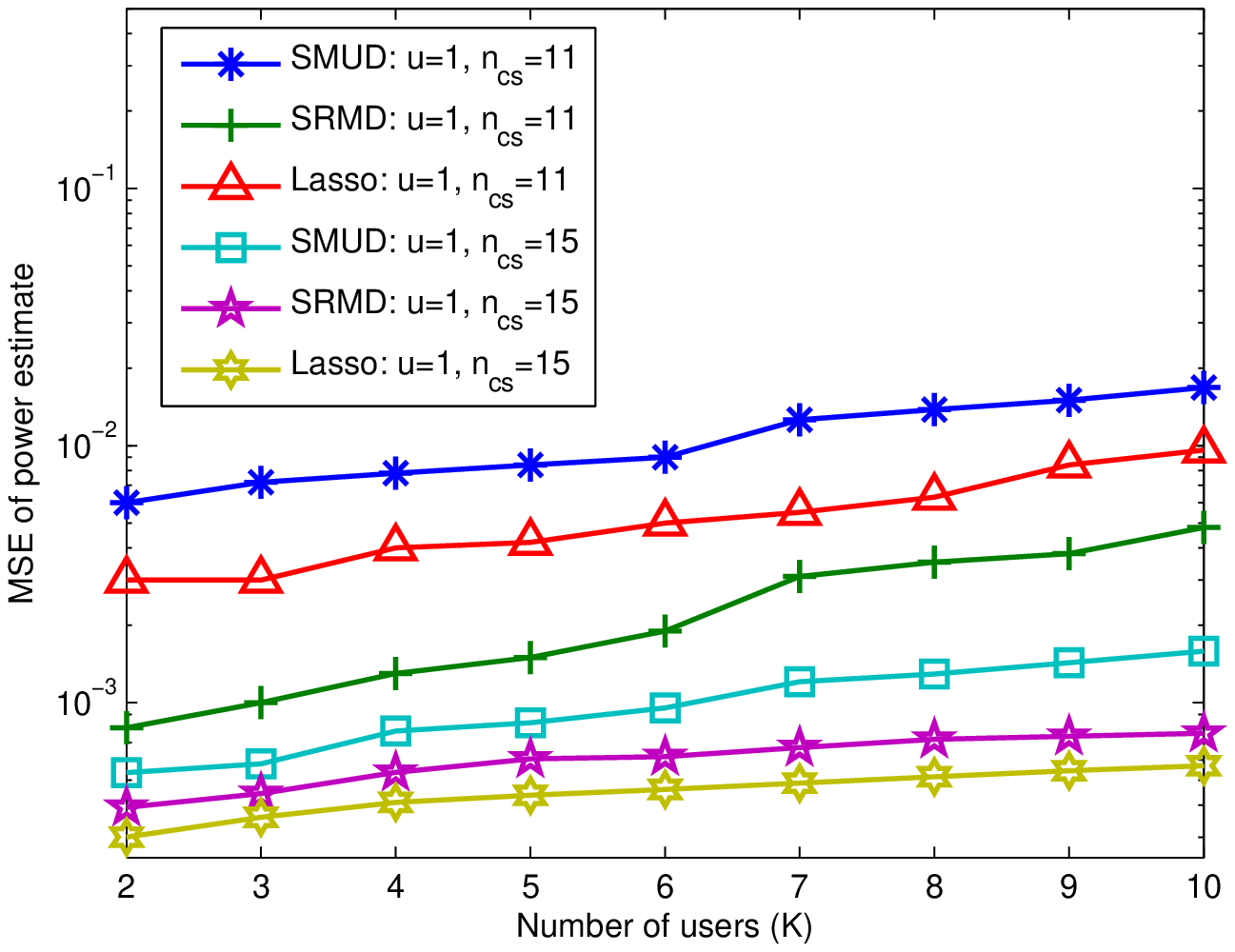}
\label{fig41}}
 \hfil \subfloat[ ]{\includegraphics[width=8.5cm]{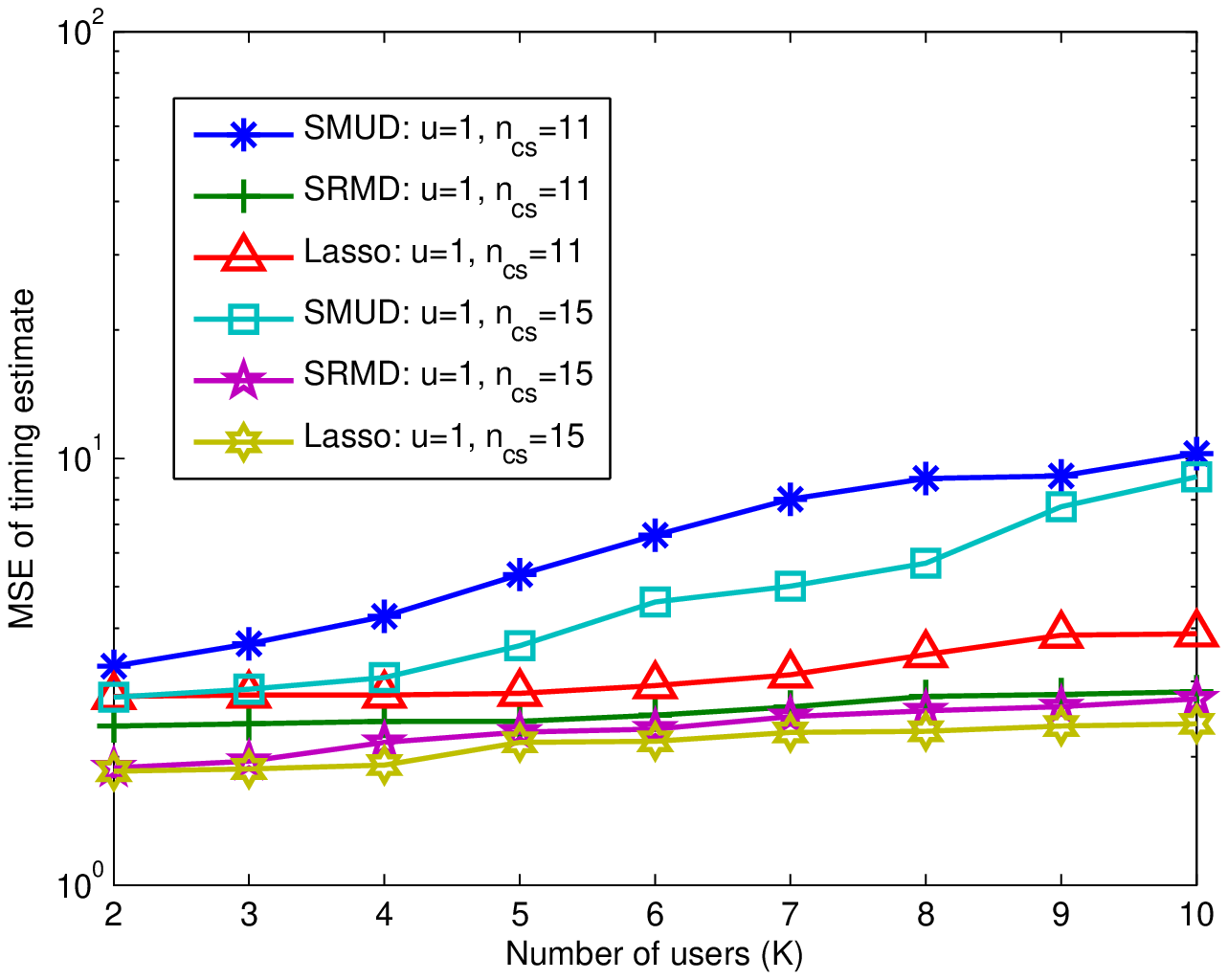}
\label{fig42}}} \caption{ MSE of initial uplink synchronization parameter estimations. (a) MSE of estimated channel power versus the number of active RTs, (b) MSE of estimated timing offset versus the number of active RTs. SNR=$10$ dB. The CRA matrix is generated with $n_{cs}=11$ and $u=1$. The CCG matrix is generated using $n_{cs}=15$ and $u=1$.  } \label{fig3}
\end{figure*}

In Figure-\ref{fig3} we plot the mean squared error (MSE) associated with the estimate 
of the power $\Gamma$ and timing offset as a function of $K$ for different code matrices. As 
expected, the MSE of power estimate by different algorithms is better with the CCG code matrix compared to the CRA matrix. It is interesting to note that the power estimation performance of the Lasso is poor for CRA matrix, however, it performs the best compared to other algorithms when we use the CCG matrix. Figure-\ref{fig3} (b) also exhibits that we can enhance the timing estimate performance of different algorithms by applying the proposed code matrix.

\section{Conclusion}
In this work, we study the dependency of performance of some state-of-the-art initial uplink synchronization algorithms on the random access (RA) code matrix which has been generated from the Zadoff-Chu (ZC) sequences. We observe that the algorithms can not perform equally for all ZC sequences. To identify the efficient ZC sequences, we apply a theory of compressive sensing. At first we develop a data model of the received signal at eNodeB over the PRACH. The data model allows us pose the IUS problem as a sparse signal representation problem on an overcomplete matrix. Consequently, we apply a sparse recovery algorithm for resolving the IUS problem. The compressive sensing theory says that the performance of sparse recovery algorithm depends on a property of the overcomplete matrix called ``mutual coherence" where a smaller value of the mutual coherence ensures better performance of the algorithm. We show that the value of mutual coherence can be controlled by properly designing the RA code matrix. We then develop a systematic procedure of code matrix design which ensures the optimum value of coherence.  The empirical results show that if we design the code matrix by using the proposed method then it also increases the performance of state of the art IUS algorithms. In particular, compared to
conventional code matrix, the IUS algorithms with the proposed code matrix show significant performance improvements in the multiuser code detection, timing offset and channel power estimations.

\appendices

\section{Proof of the Lemma-\ref{th2}}\label{app1}
We begin the proof by describing some important properties of the sinc function. Define
\begin{align}\label{eq:l}
\mathcal{S}(r)&=\left |
\frac{\mathrm{sinc} ( r M/N) }{
\mathrm{sinc} (r/N)}
\right|.
\end{align}
For any value of $|r|<\infty$,
\begin{align}\label{eq:sincp2}
 \mathcal{S}(r)&=\mathcal{S}(N-r),\\
 \label{eq:sincp3}\mathcal{S}(r)&=\mathcal{S}(r \ \mathrm{mod} \ N).
\end{align}

We have shown in Appendix-\ref{sinc} that for $\pi\sqrt{2}\le M\le N/2$, the $\mathcal{S}(r)$ also satisfies the following properties:
\begin{enumerate}
\item $\mathcal{S}(r)$ is a monotonically decreasing function for $r\in[0,1]$,
\item for $r\in[1,N-1)$, the maximizer of $\mathcal{S}(r)$ is unique and is given by 
\begin{align}\label{eq:sincp1}
\arg \max_{1\le r<N-1} \mathcal{S}(r)&=1.
\end{align}
\end{enumerate}

We are now ready to prove Lemma-\ref{th2}. For any two matrices $\bB\not=\bD$, we extend the definition of mutual coherence in \eqref{eq:coh} to ``block coherence" as
\begin{align}\label{eq:coh22}
\hat{\mu}(\bB,\bD)=\max_{j,k}\frac{|[\bB]_j^*[\bD]_k|}{\|[\bB]_j\|_2 \ \|[\bD]_k\|_2}.
\end{align}
By using the definitions of coherences in \eqref{eq:coh} and \eqref{eq:coh22}, it can be verified that
\begin{align}\label{eq:call}
\mu(\bA)=\max\{\mu(\tilde{\bE}_1),\max_{\ell\not=m}\hat{\mu}(\tilde{\bE}_{m},\tilde{\bE}_{\ell})\}.
\end{align}
Since all codes of $\bC$ are generated from the single ZC root $\tilde{u}$, by using \eqref{eq:ell} we have
\begin{align}\label{eq:ell22}
[\tilde{\bE}_{m}]_{q,k}=Z^{\tilde{u}}(q)\exp\left( \frac{-\irm 2\pi j_q}{N}\left((m-1)\frac{n_{cs} \ \tilde{u}N}{M}+(k-1)\right)\right)
\end{align}
where $[\tilde{\bE}_{m}]_{q,k}$ is the element of $\tilde{\bE}_{m}$ at its $q$-th row and $k$-th column.
Now, using \eqref{eq:coh} and \eqref{eq:ell22} we get
\begin{align}\label{eq:c1}
\mu(\tilde{\bE}_1)&=\max_{\substack{1\le k,p\le N_1\\ k\not=p}}\frac{1}{M}|[\tilde{\bE}_1]_k^*[\tilde{\bE}_1]_p|\nonumber\\
&=\frac{1}{M}|[\tilde{\bE}_1]_1^*[\tilde{\bE}_1]_2|\nonumber\\
&=\frac{1}{M}\left | \sum_{q=0}^{M-1}{\exp\left(-\irm 2 \pi q/N\right)}\right |=\mathcal{S}(1).
\end{align}
Furthermore, for any $\ell\not=m$ using \eqref{eq:coh22} we get
\begin{align}\label{eq:coh3}
\hat{\mu}(\tilde{\bE}_{m},\tilde{\bE}_{\ell})=\max_{1\le k,p\le N_1}\frac{1}{M}|[\tilde{\bE}_{m}]_k^*[\tilde{\bE}_{\ell}]_p|.
\end{align}
Now using the expression of $[\tilde{\bE}_{m}]_{q,k}$ in \eqref{eq:ell22}, for any $k,p\in\{1,2,\cdots N_1\}$ with $\ell\not=m$, we have
\begin{align}\label{eq:app1}
&\frac{1}{M}|[\tilde{\bE}_{m}]_k^*[\tilde{\bE}_{\ell}]_p|\nonumber\\
& \ \ \ = \frac{1}{M}\left |\sum_{q=0}^{M-1}{\exp\left(\frac{-\irm 2 \pi q}{N}\left(\frac{n_{cs} \ \tilde{u} N}{M}(\ell-m)+(p-k)\right)\right)}\right|\nonumber\\
%& \frac{1}{M}\left |\sum_{q=0}^{M-1}{\exp\left(-\irm 2 \pi q(\zb(s,p,m)-\zb(s,k,\ell))/N\right)}\right |\nonumber\\
& \ \ \ = \mathcal{S} \left(\frac{n_{cs} \ \tilde{u} N}{M}(\ell-m)+(p-k)\right)\nonumber\\
& \ \ \ = \mathcal{S} \left(\left(\frac{n_{cs} \ \tilde{u} N}{M}(\ell-m)+(p-k)\right) \ \mathrm{mod} \ N\right)
\end{align}
where in \eqref{eq:app1} we use the property  \eqref{eq:sincp3}. Denote
$$\xi(m,\ell,p,k)=\left(\frac{n_{cs} \ \tilde{u} N}{M}(\ell-m)+(p-k)\right) \ \mathrm{mod} \ N.$$
We see that
$$\xi(m,\ell,p,k)=N-\xi(\ell,m,k,p).$$
Hence, using \eqref{eq:sincp2} we can express \eqref{eq:app1} as
\begin{align}\label{eq:app11}
\frac{1}{M}|[\tilde{\bE}_{m}]_k^*[\tilde{\bE}_{\ell}]_p|&= \mathcal{S} \left(\min\{\xi(m,\ell,p,k),\xi(\ell,m,k,p)\}\right).
\end{align}
Note that for any values of $m,\ell,p$ and $k$ we have $\min\{\xi(m,\ell,p,k),\xi(\ell,m,k,p)\}\in[0,N/2]$. Recall that $\mathcal{S}(r)$ is a monotonically decreasing function for $0\le r\le1$. Thus, whenever $\min\{\xi(m,\ell,p,k),\xi(\ell,m,k,p)\}\le 1$, we have
\begin{align}\label{eq:t1}
&\mathcal{S} \left(\min\{\xi(m,\ell,p,k),\xi(\ell,m,k,p)\}\right)\nonumber\\
&=\mathcal{S} \left(\min\{1,\xi(m,\ell,p,k),\xi(\ell,m,k,p)\}\right)
\end{align}
 On the other hand, whenever $\min\{\xi(m,\ell,p,k),\xi(\ell,m,k,p)\}> 1$, the property of $\mathcal{S}(r)$ in \eqref{eq:sincp1} implies that
 \begin{align}\label{eq:t2}
&\mathcal{S} \left(\min\{\xi(m,\ell,p,k),\xi(\ell,m,k,p)\}\right)\nonumber\\
&<\mathcal{S} \left(\min\{1,\xi(m,\ell,p,k),\xi(\ell,m,k,p)\}\right)
\end{align}
 Now combining \eqref{eq:t1} and \eqref{eq:t2}, for any value of $\min\{\xi(m,\ell,p,k),\xi(\ell,m,k,p)\}\in[0,N/2]$ we can express \eqref{eq:app11} as
\begin{align}\label{eq:app2}
\frac{1}{M}|[\tilde{\bE}_{m}]_k^*[\tilde{\bE}_{\ell}]_p|&\le  \mathcal{S} (\min\{1,\xi(m,\ell,p,k),\xi(\ell,m,k,p)\}).
\end{align}
Consequently using \eqref{eq:coh3}, for any $\ell\not=m$ we have
\begin{align}
&\hat{\mu}(\tilde{\bE}_{m},\tilde{\bE}_{\ell})\nonumber\\
&\le \mathcal{S} \left(\min\left\{1,\min_{1\le k,p\le N_1}\{\xi(m,\ell,p,k),\xi(\ell,m,k,p)\}\right\}\right)
\end{align}
where the equality holds only if $\min_{k,p}\{\xi(m,\ell,p,k),\xi(\ell,m,k,p)\}\le 1$.
As a result, for any $\ell,m\in\{1,2,\cdots G\}$
\begin{align}\label{eq:c2}
\max_{\ell\not=m}\hat{\mu}(\tilde{\bE}_{\ell},\tilde{\bE}_{m})\le  \mathcal{S} (\min\{1,\zeta(\tilde{u})\})
\end{align}
likewise, the equality holds only if $\zeta(\tilde{u})\le 1$. Now combining \eqref{eq:c1} and \eqref{eq:c2} with \eqref{eq:call}, we obtain \eqref{eq:mu}. \feop

\section{Proof of the relation in \eqref{eq:sincp1}}\label{sinc}
By construction, $\mathcal{S}(r)$ obeys the properties of Fej\'er kernel \cite{fej}. The maximum of $\mathcal{S}(r)$ occurs at $r=0$. 
The zeros of $\mathcal{S}(r)$ are located at the non-zero multiples of $N/M$. Hence, there are $(M-1)$ number of zeros of $\mathcal{S}(r)$ on $r\in[0,N-1)$. There exists only one local maximum point between every two consecutive zeros. Consequently, $\mathcal{S}(r)$ will have $(M-2)$ number of local maxima on $r\in[0,N-1)$.

Suppose that $\pi\sqrt{2}\le M\le N/2$. For $r\in[0,N/M]$, the only zero of $\mathcal{S}(r)$ occurs at $r= N/M$  and the maximum point is at $r=0$. Hence, $\mathcal{S}(r)$ is a monotonically decreasing function for $0\le r \le N/M$. Here, $N/M\ge 2$. 
Hence
\[
\arg \max_{r \in [1, N/M] } S(r) = 1.
\]
Since $S(N-r) = S(r)$, we conclude that 
\[
\arg \max_{r \in 
[1, N/M]
\cup
[N - N/M, N-1) } S(r) = 1.
\]
We emphasize that, in above, $N$ is not included in the domain over which the maximum value is sought.
Let $r_n$ is the unique local maximum point of $\mathcal{S}(r)$ located in the interval $r_n\in (nN/M,(n+1)N/M)$. It is well known that $\mathcal{S}(r_1)\ge \mathcal{S}(r_n);  n\in\{1,2,\cdots M-2\}$  \cite{fej}. As a result, for $r\in(N/M,N-N/M)$ the maximum value of $\mathcal{S}(r)$ is $\mathcal{S}(r_1).$ Hence, to prove the relation \eqref{eq:sincp1}, it is sufficient to show that $\frac{\mathcal{S}(1)}{\mathcal{S}(r_1)}>1$ i.e., %$\mathcal{S}(1)>\mathcal{S}(r_1)$ i.e.,
\begin{align}\label{eq:bnd}
%&\frac{\mathcal{S}(1)}{\mathcal{S}(r_1)}>1\nonumber\\
&\frac{|\mathrm{sinc}(M/N)| \ |\mathrm{sinc}(r_1/N)|}{|\mathrm{sinc}(r_1 M/N)| \ |\mathrm{sinc}(1/N)|}>1.
\end{align}
Note that $r_1\in (N/M,2N/M)$. 
By differentiating $\mathrm{sinc}(r M/N)$ with respect to $r$ and equating to zero, it can be verified that the local maximum point of $|\mathrm{sinc}(r M/N)|$ for  $r\in (N/M,2N/M)$ occurs at \cite{fej2}
$$\tilde{r}=\left[3/2\pi-\frac{2}{3\pi}\right]\frac{N}{\pi M}=1.4325 \frac{N}{M}.$$
Hence,
\begin{align}\label{eq:s2}
|\mathrm{sinc}(r_1M/N)|\le|\mathrm{sinc}(\tilde{r}M/N)| =0.6824/\pi.
\end{align}
The power series expansion of $\mathrm{sinc}(x)$ is \cite{mm}
\begin{align}\label{eq:pw}
\mathrm{sinc}(x)=\sum_{n=0}^{\infty}{(-1)^n\frac{(\pi x)^{2n}}{(2n+1)!}}.
\end{align}
Now using \eqref{eq:pw}, we have
\begin{align}\label{eq:pw3}
\mathrm{sinc}(r_1/N)&=1-\frac{(\pi r_1/N)^2}{3!}+\sum_{n=2}^{\infty}{t_n}
%\mathrm{where,} \ \ t_n&=(-1)^n\frac{(\pi r_1/N)^{2n}}{(2n+1)!}.\nonumber
\end{align}
where, $t_n=(-1)^n\frac{(\pi r_1/N)^{2n}}{(2n+1)!}$. In particular, if $|\pi r_1/N|<\sqrt{48}$ then for all non-zero positive integer $k$ i.e., for $k\in\mathbb{Z}_{>0}$ we have
\begin{align}
t_{2k}+t_{2k+1}&=\frac{(\pi r_1/N)^{4k}}{(4k+1)!}-\frac{(\pi r_1/N)^{4k+2}}{(4k+3)!}\nonumber\\
&=\frac{(\pi r_1/N)^{4k}}{(4k+1)!}\left\{1-\frac{(\pi r_1/N)^{2}}{(4k+2)(4k+3)}\right\}\nonumber\\
&\ge\frac{(\pi r_1/N)^{4k}}{(4k+1)!}\left\{1-\frac{48}{6\cdot 8}\right\}=0
\end{align}
where in the last inequality we set $k=1$. Hence, $\sum_{n=2}^{\infty}{t_n}=\sum_{k=1}^{\infty}(t_{2k}+t_{2k+1})\ge 0$.
%Now using \eqref{eq:pw}, we have
%\begin{align}\label{eq:pw3}
%\mathrm{sinc}(r_1/N)&=1-\frac{(\pi r_1/N)^2}{3!}+\sum_{n=2}^{\infty}{(-1)^n\frac{(\pi r_1/N)^{2n}}{(2n+1)!}}.
%\end{align}
%Using simple mathematics, it can be shown that the last term in right hand side of \eqref{eq:pw3} will be a positive whenever $|\pi r_1/N|<6$. 
Since $r_1\in (N/M,2N/M)$ and $M\ge \pi\sqrt{2}$, hence $\pi r_1/N< 2\pi/M\le \sqrt{2}$. 
%As a result, the last term in right hand side of \eqref{eq:pw3} will be a positive quantity\footnote{In fact, using simple mathematics it can be shown that $\sum_{n=2}^{\infty}{(-1)^n\frac{(x)^{2n}}{(2n+1)!}}$ will be positive whenever $x<6$.}. 
Now from \eqref{eq:pw3}, we get
\begin{align}\label{eq:s3}
|\mathrm{sinc}(r_1/N)|&> 1-\frac{(\pi r_1/N)^2}{6}>2/3.
%&> 1-\frac{(2 \pi /M)^2}{6}
\end{align}
Since $N/M\ge2$, hence $\pi M/N\le\pi/2$. Then using a similar procedure of \eqref{eq:pw3}, we get 
\begin{align}\label{eq:temp1}
|\mathrm{sinc}(M/N)|&> 1-\frac{(\pi M)^2}{6N^2}\ge 1-\frac{\pi^2}{24}. 
%&> \mathcal{S}(r_1).
\end{align}
%where \eqref{eq:temp1} uses the fact that $\pi M/N\le\pi/2$.
Now combining \eqref{eq:s2}, \eqref{eq:s3}, \eqref{eq:temp1} and the fact $|\mathrm{sinc}(1/N)|\le 1$, we obtain the bound in \eqref{eq:bnd}.
\section{Proof of Proposition-\ref{pro1}}\label{app3}

Using the expression of $[\tilde{\bE}_{\ell}]_{n,k}$ in \eqref{eq:ell22}, we see that
%\begin{align}
%[\tilde{\bE}_{\ell}]_{q,k}=Z^r(q)\exp\left( \frac{-\irm 2\pi j_q}{M}\left(n_{cs} \ r(\ell-1)+(k-1)\frac{M}{N}\right)\right)
%\end{align}
%where $[\tilde{\bE}_{\ell}]_{q,k}$ is the element of $[\tilde{\bE}_{\ell}]$ at its $q$-th row and $k$-th column.
%Thus, 
for any $k,p\in\{1,2,\cdots N_1\}$ with $\ell\in\mI_{u1}; \ m\in\mI_{u2}$ and $u1\not=u2$, we have
\begin{align}\label{eq:app13}
&\frac{1}{M}|[\tilde{\bE}_{\ell}]_k^*[\tilde{\bE}_{m}]_p|\nonumber\\
&= \frac{1}{M}\Bigg |\sum_{n=0}^{M-1}{\exp\bigg( \frac{-\irm\pi}{M}\Big[(u2-u1)(n^2+n)}\nonumber\\
& \ \ \ \ +2n \left\{n_{cs}\left(u2(m-1)-u1(\ell-1)\right)+(p-k)\frac{M}{N}\right\}\Big] \bigg)\Bigg|\nonumber\\
%&= \frac{1}{M}\left |\sum_{n=0}^{M-1}{\exp\left(-\irm \frac{\pi}{M}(q-r)\left[n(n+1)+2n \frac{1}{(q-r)}\left\{n_{cs}\left(q(m_q-1)-r(\ell_r-1)\right)+(p-k)\frac{M}{N}\right\}\right] \right)}\right |\nonumber\\
&= \frac{1}{M}\Bigg |\sum_{n=0}^{M-1}{\exp\bigg(-\irm \frac{\pi}{M}(u2-u1)\Big[\left\{n+\vartheta(u1,u2,\ell,m,p,k)\right\}^2}\nonumber\\
& \ \ \ \ \ \ -\vartheta(u1,u2,\ell,m,p,k)^2\Big]\bigg)\Bigg |\nonumber\\
&= \frac{1}{M}\Bigg |\exp\left(\irm \frac{\pi}{M}(u2-u1)\vartheta(u1,u2,\ell,m,p,k)^2\right)\nonumber\\
& \ \ \ \ \ \ \sum_{n=0}^{M-1}{\exp\left(-\irm \frac{\pi}{M}(u2-u1)\left[\left(n+\vartheta(u1,u2,\ell,m,p,k)\right)^2\right]\right)}\Bigg |\nonumber
\end{align}
which implies \eqref{eq:elm}.

\bibliographystyle{IEEEtran}
\bibliography{rangj,Parj,Tcsb}

\end{document}